\documentclass[twocolumn]{autart} 
\usepackage{generic}
\usepackage{cite}
\usepackage{amsmath,amssymb,amsfonts}
\usepackage{subcaption}
\usepackage{algorithmic}
\usepackage{empheq}

\usepackage{algorithm}
\usepackage{graphicx}
\usepackage{textcomp}
\usepackage{cite}
\usepackage{textcomp}
\usepackage{setspace}
\usepackage{multirow}
\usepackage{xcolor}
\usepackage{bbding}
\usepackage{booktabs}
\usepackage{threeparttable}
\usepackage{bm}
\usepackage{makecell}
\usepackage{tablefootnote}
\def\BibTeX{{\rm B\kern-.05em{\sc i\kern-.025em b}\kern-.08em
 T\kern-.1667em\lower.7ex\hbox{E}\kern-.125emX}}

\usepackage{threeparttable}
\allowdisplaybreaks
\begin{document}
\begin{frontmatter}
\title{Safe and Stable Filter Design Using a Relaxed Compatibitlity Control Barrier -- Lyapunov Condition}
\thanks[footnoteinfo]{Part of the results of this manuscript has been presented as an extended abstract at the International Conference on Control 2024 \cite{ukacc}. Here we significantly extend the conference version by additionally providing a filter formulation for safe and stable controller synthesis in Section \ref{sec:ourfilter}, proposing a CBF/CLF design algorithm in Section \ref{sec:synthesis}, and conducting more comparative numerical studies in Section \ref{sec:simulation}.}
\author{Han Wang}\ead{han.wang@eng.ox.ac.uk},  
\author{Kostas Margellos}\ead{kostas.margellos@eng.ox.ac.uk}, 
\author{Antonis Papachristodoulou}\ead{antonis@eng.ox.ac.uk}        
\address{OX1 3PJ, Department of Engineering Science, University of Oxford, Oxford, United Kingdom.} 

\maketitle

\begin{abstract}
In this paper, we propose a quadratic programming-based filter for safe and stable controller design, via a Control Barrier Function (CBF) and a Control Lyapunov Function (CLF). Our method guarantees safety and local asymptotic stability without the need for an asymptotically stabilizing control law. Feasibility of the proposed program is ensured under a mild regularity condition, termed \emph{relaxed compatibility} between the CLF and CBF. The resulting optimal control law is guaranteed to be locally Lipschitz continuous. We also analyze the closed-loop behaviour by characterizing the equilibrium points, and verifying that there are no equilibrium points in the interior of the control invariant set except at the origin. For a polynomial system and a semi-algebraic safe set, we provide a sum-of-squares program to design a relaxed compatible pair of CLF and CBF. The proposed approach is compared with other methods in the literature using numerical examples, exhibits superior filter performance and guarantees safety and local stability.
\end{abstract}
\end{frontmatter}

\section{Introduction}
\label{sec:introduction}
Safety and stability are essential properties for modern automation applications \cite{alleyne2023control}. In control theory, one popular methodology to verify and enforce these properties is to use \emph{certificate functions}. For a nonlinear autonomous system, stability can be verified by constructing a Lyapunov function \cite{khalil2009lyapunov}. When a control input is considered, the concept of a Control Lyapunov Function (CLF) has been introduced\cite{artstein1983stabilization}. With a Control Lyapunov Function in hand, a stabilizing controller can be designed using Sontag's universal formula \cite{sontag1989smooth}. On the other hand, safety verification can be achieved by defining a barrier function over the state-space \cite{prajna2004safety}. The zero-level set of a barrier function is forward invariant and discriminates the safe and unsafe regions. Analogous to CLFs, Control Barrier Functions (CBFs) have been introduced to guide controller design for safety \cite{wieland2007constructive,ames2016control}.

To guarantee stability and safety simultaneously for a dynamical system, a quadratic programming-based filter has been proposed \cite{ames2016control}. CLF and CBF constitute separate constraints in such a filter for stability and safety. The filter can accommodate any locally Lipschitz continuous reference signal and subsequently provides the closest certified control signal. For nonlinear control affine systems, CLF and CBF constraints are linear in the input for any given state. Feasibility, however, is not guaranteed: as a trade-off, the CLF constraint is relaxed using a slack variable in the program. Consequently, local stability for the closed-loop system using the designed controller is not guaranteed \cite{reis2020control}. Additionally, equilibrium points, other than the origin, exist. 
\begin{table*}[ht]
\normalsize
\caption{Comparisons on closed-loop behaviours of our filter and the other CLF-CBF based filters in the literature. By $\mathcal{O}_l$ we mean the $l$-sublevel set of a CLF, $\mathcal{O}_{\eta\le l}$ denotes the $\eta$-sublevel set of the CLF, where $0\le n\le l$. It holds that $\mathcal{O}_{\eta\le l}\subseteq \mathcal{O}_{l}$. Note that in \cite{schneeberger2024advanced} compatibility between CBF and CLF is strict (in a sense specified formally in the sequel), however, linear independence of the associated constraints is imposed as an assumption.}
\begin{tabular}{ccccccc}
\Xhline{3\arrayrulewidth}
Paper & Local Stability & Interior Equilibria & ROA & Stabilizing Controller & Compatibility & CBF/CLF Design \\ \hline
This Work              & $\checkmark$   & Do not Exist & $\mathcal{O}_l$   & Not Required & Relaxed        & $\checkmark$   \\ 
\cite{ames2016control}         & $-$ & Exist     & $-$ & Not Required & Not Required & $-$ \\ 
\cite{jankovic2018robust}           & $\checkmark$   & Exist     & $-$ & Not Required & Not Required & $-$ \\ 
\cite{tan2021undesired}      & $\checkmark$   & Do not Exist & $\mathcal{O}_{l}$   & Required     & Not Required & $-$ \\ 
\cite{cortez2022compatibility} & $\checkmark$   &  Do not Exist  & $\mathcal{O}_{\eta\le l}$   & Required     & Strict       & $-$ \\ 
\cite{mestres2022optimization}           & $\checkmark$   &  Do not Exist  & $\mathcal{O}_{\eta\le l}$   & Not Required & Strict       & $-$ \\ 
\cite{schneeberger2024advanced} & $\checkmark$ & Do not Exist & $\mathcal{O}_{l}$ & Not Required & Strict & $\checkmark$\\
\Xhline{3\arrayrulewidth}
\end{tabular}
\label{tab:my-table}
\end{table*}
To address these issues, many improved filters have been proposed. The authors in \cite{romdlony2016stabilization} unify the CLF and CBF into one function called CLBF for simultaneous stability and safety, but only for a bounded control invariant set. Given a CLF and a CBF, a CLBF can be constructed through a linear combination of these functions. In \cite{jankovic2018robust}, the CLF constraint is modified with additional parameters. By properly designing these parameters, the closed-loop system can be locally stable at the origin. If the reference control signal stabilizes the system, the filter \cite{ames2016control} is shown to guarantee local stability \cite{tan2021undesired}. Under a similar assumption, \cite{cortez2022compatibility} proposes a new filter that only has a CBF constraint. Sufficient conditions to estimate the region of attraction (ROA) are also provided. To better accommodate the CLF and CBF constraints, \cite{mestres2022optimization} proposes to lift either the CLF or the CBF constraint into the objective function. When the CLF constraint is lifted as a penalty term, the closed-loop system is shown to be locally stable in a non-empty region of attraction (ROA) if the penalty parameter is larger than a certain level. 

In \cite{ames2016control}, the CLF and CBF are obtained separately. The potential conflict between these two functions is the main reason for the undesired closed-loop behaviour. The concept of \emph{compatibility} has been identified as a key property to efficiently accommodate the conflict \cite{cortez2022compatibility,mestres2022optimization}. Compatibility is also related to the control sharing property \cite{xu2018constrained}. In essence, this property necessitates the existence of a controller that satisfies both the CLF and CBF constraints at every state. To design compatible CLF and CBF, a sum-of-squares programming-based synthesis method has been proposed \cite{schneeberger2023sos}. However, it is important to note that achieving compatibility for the entire state space may be impossible if the complementary set of the control invariant set is bounded \cite{braun2020comment}.

In this work, we propose a new filter based on the \emph{relaxed compatibility} condition for a CBF and CLF. We demonstrate that our method obtains the desired closed-loop behaviour, i.e. local asymptotic stability and elimination of interior equilibrium points. Our method does not require a stabilizing nominal controller to enhance stability of the designed optimal controller. Moreover, the optimal controller is guaranteed to be locally Lipschitz continuous inside the invariant set without an \textit{a priori} assumption on the linear independence of the CBF- and CLF- induced linear constraints, which is assumed in \cite{schneeberger2024advanced}. Additionally, we provide a design method for a relaxed compatible pair of CLF and CBF for a polynomial dynamical system using sum-of-squares programming. The inherent nonlinearities in the program are addressed through an iterative algorithm. We validate the efficacy of our method by comparative studies against other stat-of-the-art methods, and our method shows superior filter performance.

A classification of the most closely related results in the literature and a qualitative comparison of their closed-loop behaviour with respect to the proposed approach is provided in Table \ref{tab:my-table}. Under mild conditions, i.e. with relaxed compatibility and without pre-stabilizing controller, our method guarantees local stability, eliminates all the interior equilibrium points, and results in a large ROA $\mathcal{O}_l$, where $l$ denotes the $l$-superlevel set of a CLF. Numerical comparisons of filter performance in terms of optimality and CBF/CLF design with other works are illustrated by numerical studies in Section \ref{sec:simulation}.

In Section \ref{sec:preliminary} we provide some mathematical preliminaries. Our filter is proposed and analyzed in Section \ref{sec:ourfilter}. The CBF/CLF design program is presented in Section \ref{sec:synthesis}. In Section \ref{sec:simulation}, we conduct comparative numerical studies, while Section \ref{sec:conclusion} provides some concluding remarks.

\section{Preliminaries}
\label{sec:preliminary}

\subsection{Notation}
$\mathbb{R}^n$ denotes the set of real vectors with dimension $n$. $\mathbb{R}[x]$ denotes the set of real polynomials in $x$. $\Sigma[x]$ denotes the set of sum-of-squares polynomials in $x$. For a vector $y$, $||y||$ denotes its $2$-norm. For a set $\mathcal{C}$, $\mathrm{Int}(\mathcal{C})$ and $\mathrm{cl}(\mathrm{C})$ denote its interior set and closure, respectively. For a closed set $\mathcal{D}$, $\partial \mathcal{D}$ denotes its boundary. For non-zero vectors $a\in\mathbb{R}^n,b\in\mathbb{R}^n$, $a//b$ indicates they are parallel. A continuous function $\alpha(\cdot):[-c,d]\to(-\infty,\infty)$ for $c,d>0$ is said to belong to extended class-$\mathcal{K}$ if it is strictly increasing, and $\alpha(0)=0$. A continuous function $\gamma(\cdot):\mathbb{R}^n\to\mathbb{R}$ is said to be positive definite if $\gamma(0)=0$, and $\gamma(x)>0,\forall x\ne 0$. For a polynomial $p(x)$, $\mathrm{deg}(p(x))$ represents its degree.
\subsection{Control Barrier \& Lyapunov Functions}

Consider a nonlinear control-affine system 
\begin{equation}\label{eq:system}
    \dot x = f(x)+g(x)u,
\end{equation}
where $x(t)\in\mathcal{X}\subset\mathbb{R}^n$ denotes the state of the system and $u(t)\in\mathbb{R}^m$ denotes the input. Here $f(x):\mathcal{X}\to \mathbb{R}^n$, $g(x):\mathcal{X}\to\mathbb{R}^m$ are {locally }Lipschitz continuous functions in $\mathcal{X}$, and $f(0)=0$. Our goal is to design a locally Lispchitz continuous state feedback controller $u(\cdot):~ \mathcal{X} \to \mathbb{R}^m$ such that the solution of the closed-loop system $\dot x=f(x)+g(x)u(x)$ that starts from $x(0)=x_0$, stays within a \emph{safe set} $\mathcal{S}$ for every $t$ that belongs to the time-domain over which solutions are defined. If such a controller $u(\cdot)$ exists, we say that the system is \emph{safe}. 
 
\begin{defn}[\protect{\cite[Definition 5]{ames2016control}}]\label{def:CBF}
    Consider system \eqref{eq:system}, and a safe set $\mathcal{S}$. A differentiable function $b(x):{\mathcal{X}\to\mathbb{R}}$ is called a \emph{Control Barrier Function (CBF)} if $\mathcal{B}:=\{x\in\mathbb{R}^n:b(x)\ge 0\}\subseteq \mathcal{S}${, and there exists a class-$\mathcal{K}$ function $\alpha(\cdot)$, and a locally Lipschitz continuous state-feedback controller with $u(x):\mathcal{X}\to\mathbb{R}^m$}, such that
        \begin{equation}\label{eq:cbfcondition}
        \mathcal{L}_fb(x)+\mathcal{L}_gb(x)u(x)+\alpha(b(x))\ge 0, \forall x\in\mathcal{X},
        \end{equation}
    where $\mathcal{L}_fb(x)=\frac{\partial b(x)}{\partial x}f(x)$, and $\mathcal{L}_gb(x)=\frac{\partial b(x)}{\partial x}g(x)$.
\end{defn}

\begin{defn}\label{def:CLF}
    A locally positive definite and differentiable function $V(\cdot):\mathcal{X}\to \mathbb{R}$ is called a \emph{Control Lyapunov Function (CLF)} if there exist a positive definite function $\gamma(\cdot):\mathbb{R}^n\to \mathbb{R}$, and a locally Lispchitz continuous state-feedback controller with $u(x):\mathcal{X}\to\mathbb{R}^m$, such that     \begin{align}
\mathcal{L}_fV(x)+\mathcal{L}_gV(x)u(x)+\gamma(x)\le 0,\forall x\in\mathcal{X},\label{eq:clfflow}
    \end{align}
    where $\mathcal{L}_fV(x)=\frac{\partial V(x)}{\partial x}f(x)$, $\mathcal{L}_gV(x)=\frac{\partial V(x)}{\partial x}g(x)$.
\end{defn}

The definition below is known as a \emph{compatibility} condition \cite{cortez2022compatibility,mestres2022optimization,schneeberger2023sos}, and captures the cases under which the CLF and CBF conditions are both feasible. 

\begin{defn}[Compatibility\protect{\cite[Definition 1]{schneeberger2023sos}}]\label{def:compatible}
    Consider system \eqref{eq:system}, CLF $V(x)$, CBF $b(x)$, extended class-$\mathcal{K}$ function $\alpha(\cdot):\mathbb{R}\to \mathbb{R}$, and positive definite function $\gamma(\cdot):\mathbb{R}^n\to \mathbb{R}$. $V(x)$ and $b(x)$ are said to be \emph{compatible} if there exists a locally Lipschitz continuous controller $u(x):\mathcal{X}\to\mathbb{R}^m$, such that for any $x\in\mathcal{X}$:
    \begin{equation}\label{eq:def6eq}
        \begin{split}
            &\mathcal{L}_fb(x)+\mathcal{L}_gb(x)u(x)+\alpha(b(x))\ge 0,\\
            &\mathcal{L}_fV(x)+\mathcal{L}_gV(x)u(x)+\gamma(x)\le 0.
        \end{split}
    \end{equation}
\end{defn}
In the next section we propose a relaxed version of this condition, which is at the core of our analysis.

\subsection{Sum-of-Squares Programming}
\begin{defn}\label{def:sos}
A polynomial $p(x)$ is said to be a sum-of-squares polynomial in $x\in\mathbb{R}^n$ if there exist $M$ polynomials $p_i(x)$, {$i=1,\ldots,M$,} such that 
\begin{equation}\label{eq:sosdecomposition}
    p(x)=\sum_{i=1}^Mp_i(x)^2.
\end{equation}
\end{defn}
We also call \eqref{eq:sosdecomposition} a sum-of-squares decomposition for $p(x)$. Clearly, if a function $p(x)$ has a sum-of-squares decomposition, then it is non-negative for all $x\in\mathbb{R}^n$. Computing the sum-of-squares decomposition \eqref{eq:sosdecomposition} is equivalent to a positive semidefinite feasibility program, which is a convex optimization program.
\begin{lem}\label{lem:sos}
Consider a polynomial $p(x)$ of degree $2d$ in $x\in\mathbb{R}^n$. Let $z(x)$ be a vector of all monomials of degree less than or equal to $d$. Then $p(x)$ admits a sum-of-squares decomposition if and only if
\begin{equation}\label{eq:sossdp}
    p(x) = z(x)^\top Q z(x), Q \succeq 0.
\end{equation}
\end{lem}
In Lemma \ref{lem:sos}, $z(x)$ is a user-defined monomial basis if $d$ and $n$ are fixed. In the worst case, $z(x)$ has $\left( \begin{array}{c}
n + d\\
d
\end{array} \right)$ components, and $Q$ is a $\left( \begin{array}{c}
n + d\\
d
\end{array} \right) \times \left( \begin{array}{c}
n + d\\
d
\end{array} \right)$ square matrix. The necessity of Lemma \ref{lem:sos} is natural from the definition of a positive semi-definite matrix, considering the monomial $z(x)$ as a vector of new variables whose $i$-th element is $z_i$. The sufficiency is shown by factorizing $Q=L^\top L$. Then $z(x)^\top Q z(x)=(Lz(x))^\top Lz(x)=||Lz(x)||^2\ge 0$.

Given $z(x)$, finding $Q$ to decompose $f(x)$ following \eqref{eq:sossdp} is a semi-definite program, which can be solved efficiently using interior point methods. Selecting the basis $z(x)$ depends on the structure of $p(x)$ to be decomposed. 
\begin{defn}\label{def:semi-algebrai}
A set $\mathcal{X}\subset \mathbb{R}^n$ is \emph{semi-algebraic} if it can be represented by polynomial equality and inequality constraints. If there are only equality constraints, the set is called \emph{algebraic}.
\end{defn}

\begin{lem}[Lossless S-lemma]\label{lem:S-procedure}
Suppose $t_i(x)\in\Sigma[x],i\in\mathcal{I}$, then
\begin{align*}
    &p(x)-\sum_{i\in\mathcal{I}}t_i(x)q_i(x)\in\Sigma[x]\Rightarrow \\
    &p(x)\ge 0,~ \forall x\in\bigcap_{i\in\mathcal{I}}\{x|q_i(x)\ge 0\}.
\end{align*}

Suppose $l_i(x)\in\mathbb{R}[x],i\in\mathcal{I}$, then
\begin{align*}
    &p(x)-\sum_{i\in\mathcal{I}}l_i(x)q_i(x)\in\Sigma[x]\Rightarrow \\
    &p(x)\ge 0,~ \forall x\in\bigcap_{i\in\mathcal{I}}\{x|q_i(x)= 0\}.
\end{align*}
\end{lem}

\section{Safety and Stability Filter}
\label{sec:ourfilter}
\subsection{Problem statement and motivating example}
Consider system \eqref{eq:system}, and a locally Lipschitz continuous controller $\pi:\mathcal{X}\to \mathbb{R}^m$ satisfying $\pi(0)=0$. Our goal is to design a locally Lipschitz continuous controller $u^*(x)$ that locally stabilizes system \eqref{eq:system} around the origin, and guarantees forward invariance of the set $\mathcal{B}=\{x\in\mathbb{R}^n:b(x)\ge 0\}$. Moreover, the optimal controller $u^*(x)$ should be close to $\pi(x)$ as much as possible.
For any $x \in \mathcal{X}$, the proposed safety and stability filter takes the form 
\begin{equation}\label{eq:ourfilter}
    \begin{split}
        (u^*(x),s^*(x))&=\mathop{\arg\min}_{u\in\mathbb{R}^m,s\in\mathbb{R}}~\frac{1}{2}||u-\pi(x)||^2+\frac{p}{2}(s-1)^2\\
        \mathrm{subject~to}\\
        F_1(u,s)=\mathcal{L}_f&b(x)+\mathcal{L}_gb(x)u+s(\alpha(b(x)))\ge 0,\\
        F_2(u)=\mathcal{L}_f&V(x)+\mathcal{L}_gV(x)u+\beta(b(x))\gamma(x) \le 0,
    \end{split}
\end{equation}
where $\alpha(\cdot):\mathbb{R}^n\to \mathbb{R}$ and $\beta(\cdot):\mathbb{R}^n\to \mathbb{R}$ are extended class-$\mathcal{K}$ functions, $\gamma(\cdot):\mathbb{R}^n\to\mathbb{R}$ is a positive definite function. In addition, $\beta(x)\le 1$, for any $x \in \mathcal{X}$. One typical choice for $\beta(\cdot)$ is $\tanh(\cdot)$. The filter is constructed at the current state $x$ given the value of $\pi(x)$. In this way the optimal solution $u^*$ and $s^*$ are both functions of $x$. In particular, we have that $u^*(x):\mathcal{X}\to \mathbb{R}^m$, and $s^*(x)\in\mathcal{X}\to\mathbb{R}$ since we only consider one CBF constraint. Unlike the filters proposed in \cite{ames2016control}, \cite{tan2021undesired} which relax the CLF constraint with a slack variable, our proposed filter \eqref{eq:ourfilter} uses an adaptive variable $s$ to ensure feasibility while promoting safety, and a term $\beta(b(x))\gamma(x)$ to ensure feasibility while promoting stability. 

{The intuition for $\beta(b(x))$ is that, for $x\in\{x\in\mathbb{R}^n:b(x)>0\}$, we have that $\beta(b(x))>0$ (since $\beta$ is an extended class-$\mathcal{K}$ function), and also $\dot V<0$ holds. Moreover, for larger $b(x)$, $\beta(b(x))$ accelerates the convergence speed by amplifying $\beta(b(x))\gamma(x)$. For any $x$ such that $\beta(b(x))$ takes the largest admissible value $1$, our adapted CLF constraint is equivalent to the original one. This happens when $b(x)$ (the argument of $\beta(\cdot)$) is large enough, i.e., the system is safe to a certain level. $\beta(x)$ can be thought of as an activation function to trigger the CLF constraint, based on the CBF constraint. 

Prior to analyzing our proposed filter, to
better understand the benefits of introducing $s$ and $\beta(\cdot)$ in terms of feasibility, safety, and stability, we provide the following motivating example.}

{
\begin{exmp}\label{ex:2}
    Consider a second-order linear system
\begin{equation}\label{eq:benchmark}
    \dot x = x+u,
\end{equation}
where $x=[x_1,x_2]^\top$, $u=[u_1,u_2]^\top$. A bounded obstacle is defined by the set $\{x\in\mathbb{R}^2:||x-(0,4)||^2\le 2\}$. This benchmark case has been considered in \cite{tan2021undesired,reis2020control,mestres2022optimization}. In these papers, a candidate CLF is $V(x)=x^\top x$, and a CBF 
\[
b(x)=||x-(0,4)||^2-4,
\]
while $\gamma(x)=V(x)$, $\alpha(b(x))=b(x)$. 
The original CBF constraint is given by $\dot b(x)+b(x)\ge 0$, which is
        $\begin{bmatrix}
            2x_1\\2x_2-8
        \end{bmatrix}^\top
        \begin{bmatrix}
            u_1\\u_2
        \end{bmatrix}
        +3x_1^2+3x_2^2-16x_2+12\ge 0.$
    The original CLF constraint is given by $\dot V(x)+V(x)\le 0$, which is$
        \begin{bmatrix}
            2x_1\\2x_2
        \end{bmatrix}^\top
        \begin{bmatrix}
            u_1\\u_2
        \end{bmatrix}
        +3x_1^2+3x_2^2\ge 0.
    $
Given a state $x$, both constraints are affine in $u$, thus define two half planes on $\mathbb{R}^2$. When \[\begin{bmatrix}
    2x_1\\2x_2-8
\end{bmatrix}//
\begin{bmatrix}
    2x_1\\2x_2
\end{bmatrix},\] the intersection of the two half planes is potentially empty. Clearly, this happens only when $x_1=0$. Under this, the CBF and CLF constraints are then given by
$2x_2u_2+3x_2^2\le 0,(2x_2-8)u_2+3x_2^2-16x_2+12\ge 0.$
It is easy to verify that, when $x_2\le 4$, there always exist $u_2$ that satisfies these constraints. Considering $x_2>4$, we have$
-\frac{3x_2^2-16x_2+12}{2x_2-8}\le u_2\le -\frac{3x_2}{2}.$
However, the simultaneous satisfaction of these inequalities is impossible if 
\[-\frac{3x_2^2-16x_2+12}{2x_2-8}>-\frac{3x_2}{2}\Longleftrightarrow x_2>3.\] Therefore, for any $x_1=0,x_2>4$, there does \emph{not} exist $u(x)$ that satisfies the CLF and CBF constraints, and the CLF $V(x)$ and CBF $b(x)$ are not compatible according to Definition \ref{def:compatible}. 

Now consider the constraint $F_1(u,s)\ge 0$ in \eqref{eq:ourfilter} given by
\begin{align*}
            2x_1u_1+(2x_2-8)u_2+(x_1^2+x_2^2&-8x_2+12)s\\
            &+2x_1^2+2x_2^2-8x_2\ge 0,
\end{align*}
and the CLF constraint $F_2(u)\le 0$ in \eqref{eq:ourfilter}, given by
\[  2x_1u_1+2x_2u_2+2x_1^2+2x_2^2+\beta(b(x))(x_1^2+x_2^2)\le 0.\]
Given a state $x$, both constraints are affine in $u$ and $s$, thus define two half planes on $\mathbb{R}^3$. The intersection of the two half planes is potentially empty if
\[
\begin{bmatrix}
            2x_1\\2x_2-8\\x_1^2+x_2^2-8x_2+12
        \end{bmatrix}//\begin{bmatrix}
    2x_1\\2x_2\\0
\end{bmatrix}.
\]
This happens when $x_1=0$ and $x_1^2+x_2^2-8x_2+12=0$, which imply $x_1=0,x_2=2$ or $x_1=0,x_2=6$. These two points both lie on the boundary of $\{x\in\mathbb{R}^2:b(x)=0\}$. Now it is clear that the additional decision variable, $s$, enlarges the affine constraint space at any $x\notin\{x\in\mathbb{R}^2:b(x)=0\}$, thus improving feasibility. On the set $\{x\in\mathbb{R}^2:b(x)=0\}$, we clearly have that constraint $F_1(u,s)\ge 0$ is the same as the nominal CBF constraint, i.e. the one without $s$.

Considering $x_1=0$, $x_2=2$, the adapted CBF constraint $F_1(u,s)\ge 0$ implies $u_2\le -2$, the adapted CLF constraint is
\[4{u_2} + 8 + 4\underbrace {\beta (b(x))}_{b(x) = 0} \le 0\Longleftrightarrow u_2\le -2.\]
At this point, any $u_1\in\mathbb{R}$ and $u_2\le -2$ satisfy the two constraints.
Considering $x_1=0,x_2=6$, the adapted CBF constraint implies $u_2\ge -6$, the adapted CLF constraint is
\[12{u_2} + 72 + 4\underbrace {\beta (b(x))}_{b(x) = 0} \le 0\Longleftrightarrow u_2\le -6.\]
Similarly, at this point, any $u_1\in\mathbb{R}$, and $u_2=-6$ satisfy the CBF and CLF constraints simultaneously. Recall that at $x_1=0$, $x_2=6>4$, there exists no $u(x)$ that satisfies the CBF and CLF constraints (as required by the standard compatibility definition). This illustrates the potential benefit of introducing $\beta(b(x))$, as it promotes feasibility on the set $\{x\in\mathbb{R}^2:b(x)=0\}$.

\end{exmp}
}

{
Example \ref{ex:2} shows that our filter \eqref{eq:ourfilter} introduces $s$ and $\beta(b(x))$ to adapt the original CBF/CLF constraints to guarantee feasibility. What remains to answer is whether the adapted CBF/CLF constraints are feasible on $\{x\in\mathbb{R}^n:b(x)=0\}$. This motivates the concept of \emph{relaxed compatibility}.
}

\begin{defn}[Relaxed Compatibility]\label{def:compalyapunov}
    Consider system \eqref{eq:system}, CLF $V(x)$, and CBF $b(x)$. $V(x)$ and $b(x)$ are said to satisfy the relaxed compatibility condition if there exists a locally Lipschitz continuous $u(x):\mathcal{X}\to\mathbb{R}^m$, such that for any $x\in\{x\in\mathbb{R}^n:b(x)=0\}$:
\begin{align}
    \mathcal{L}_fb(x)+\mathcal{L}_gb(x)u(x)&\ge 0,\quad \nonumber \\
    \mathcal{L}_fV(x)+\mathcal{L}_gV(x)u(x)&\le 0. \label{eq:def3eq2}
\end{align}
\end{defn}

Compatibility is naturally sufficient but not necessary for relaxed compatibility. There are two main differences between them:
\begin{enumerate}
    \item Conditions for relaxed compatibility only require the existence of a controller $u(x)$ that satisfies the CBF constraint and the relaxed CLF constraint ($\dot V(x)\ge 0$) for any $x\in\partial \mathcal{B}$, whereas the existence of controller is required for any $x\in\mathbb{R}^n$ for compatibility. 
    \item The CLF constraint in \eqref{eq:def3eq2} is non-strict and does not include the convergence rate term $\gamma(x)$ on $\partial\mathcal{B}$, whilst is strict in the sense of $\dot V<0,\forall x\ne0$ in \eqref{eq:def6eq}. 
\end{enumerate}
Relaxed compatibility refers to compatibility of \emph{invariance} and \emph{stability} that only relies on a CBF (control invariant set), a CLF, and the system dynamics. The motivation for (2) comes from a geometric observation in \cite{braun2020comment}. Let $\mathcal{B}^c$ denote the set complement of $\mathcal{B}$, and suppose $\mathcal{B}^c$ is bounded. It is then shown that there is no locally Lipschitz continuous controller $u(x)$ that guarantees that both $\dot V <0$ and $\dot b \ge 0$ on some points on the boundary of the invariant set\cite{braun2020comment}. As a result, compatibility can not hold on these points. Relaxed compatibility therefore proposes a milder condition that $\dot V\le 0$ and $\dot b\ge 0$ on these points. 

\begin{thm}\label{th:feasibility}
    {Consider system \eqref{eq:system}. Suppose there exist a CBF $b(x)$ and a CBF $V(x)$ that satisfy the relaxed compatibility condition as per Definition \ref{def:compalyapunov}.} Then the optimization problem \eqref{eq:ourfilter} is feasible for any $x\in\mathcal{X}$.
\end{thm}

\begin{pf}
We will show that both constraints in \eqref{eq:ourfilter} are feasible.
To this end, by Definition \ref{def:CLF}, for any $x\in\mathcal{X}$, there exists $u'$, such that $\mathcal{L}_fV(x)+\mathcal{L}_gV(x)u'+\gamma(x)\le 0$. Given that $\gamma(\cdot)$ is non-negative, and $\beta(x)\le 1$, we have that for any $x\in\mathcal{X}$, $\beta(b(x))\gamma(x)\le \gamma(x)$. This indicates that for any $x\in\mathcal{X}$, 
\begin{align}
   F_2(u') &=\mathcal{L}_fV(x)+\mathcal{L}_gV(x)u'+\beta(b(x))\gamma(x) \nonumber \\
   &\le \mathcal{L}_fV(x)+\mathcal{L}_gV(x)u'+\gamma(x) \le 0,
\end{align} 
where the second inequality is due to the property of the selected $u'$.

Consider the same $u'$ for the CBF constraint, and let
\begin{equation}
    s' = \left\{ \begin{array}{l}
    -\frac{\mathcal{L}_fb(x)+\mathcal{L}_gb(x)u'}{\alpha(b(x))}+1,~\mathrm{if~}b(x)>0,
\\
-\frac{\mathcal{L}_fb(x)+\mathcal{L}_gb(x)u'}{\alpha(b(x))}-1,~\mathrm{if~}b(x)<0,\\
0,~~~~~~~~~~~~~~~~~~~~~~~~~~~~~\mathrm{if~}b(x)=0,
\end{array} \right.
\end{equation}
and notice that the division with $\alpha(b(x))$ is admissible for the values of $b(x)$ in these cases.
We have that for any $x\in\{x\in\mathbb{R}^n:b(x)\ne 0\}$,  
\begin{align}
    F_1(u',s')=\mathcal{L}_fb(x)+\mathcal{L}_gb(x)u'+s'(\alpha(b(x)))\ge 0,
\end{align} 
where the inequality follows by substituting in place of $s'$ the expressions in the first two branches of that correspond to $x\in \mathcal{X}$ such that $b(x)\ne 0$.
Additionally, for any $x\in\mathcal{X}$ such that $b(x)=0$, by the relaxed compatibility condition \eqref{eq:def3eq2}, we directly have that $F_1(u',s') \ge 0$. As such, in any case feasibility of  \eqref{eq:ourfilter} is ensured, thus concluding the proof.\hfill\qed
\end{pf}

\subsection{Equilibrium Characterization}
Theorem \ref{th:feasibility} shows that under the relaxed compatibility condition of Definition \ref{def:compalyapunov}, \eqref{eq:ourfilter} is feasible for any $x\in\mathcal{X}$. 
We now provide a closed-form expression for the optimal solution of \eqref{eq:ourfilter}.

\begin{thm}\label{th:analytical}
    Consider system \eqref{eq:system}, and a safe set $\mathcal{S}\subset\mathbb{R}^n$. Suppose there exist a CBF $b(x)$ and a CLF $V(x)$ that satisfy the relaxed compatibility condition. The optimal controller $u^*(x)$ obtained by the safety and stability filter \eqref{eq:ourfilter} is given by
   \begin{subequations}\label{eq:controller}
    \begin{empheq}[left={u^*=\empheqlbrace\,}]{align}
       &\pi,\forall x\in\Omega_{\overline{clf}}^{\overline{cbf}}\cup \Omega_{clf,2}^{\overline{cbf}}\cup\Omega_{clf,5}^{cbf}\cup\Omega_{\overline{clf},2}^{cbf},\label{eq:control1}\\ 
      &\pi-\frac{F_b}{F'_b}\mathcal{L}_gb^\top,\forall x\in\Omega_{\overline{clf},1}^{cbf}\cup\Omega_{clf,4}^{cbf},\label{eq:control2}\\
      &\pi-\frac{F_V}{\mathcal{L}_gV\mathcal{L}_gV^\top }\mathcal{L}_gV^\top,\forall x\in\Omega_{clf,1}^{\overline{cbf}},\label{eq:control3}\\
      &\pi+\begin{bmatrix}
                \mathcal{L}_gb^\top\\-\mathcal{L}_gV^\top
            \end{bmatrix}^\top A(x)^{-1}\begin{bmatrix}
            -F_b(x)\\F_V(x)
        \end{bmatrix},\nonumber\\
       &~~~~\forall x\in\Omega_{clf,1}^{cbf}, \label{eq:control4}\\
        &\pi+\frac{F_V}{\mathcal{L}_gV\mathcal{L}_gV^\top}\mathcal{L}_gV^\top,\nonumber\\
       &~~~~\forall x\in\Omega_{clf,2}^{cbf}\cup\Omega_{clf,3}^{cbf},\label{eq:control5}
    \end{empheq}
\end{subequations}

    where 
    \[
    A(x)=\begin{bmatrix}
        \mathcal{L}_gb\mathcal{L}_gb^\top+\frac{\alpha(b(x))^2}{p} & -\mathcal{L}_gb\mathcal{L}_gV^\top \\
        \mathcal{L}_gV\mathcal{L}_gb^\top & -\mathcal{L}_gV\mathcal{L}_gV^\top 
    \end{bmatrix},
    \]
$F_b(x)=\mathcal{L}_gb(x)\pi(x)+\mathcal{L}_fb(x)+\alpha(b(x))$, $F'_b(x)=\mathcal{L}_gb(x)\mathcal{L}_gb(x)^\top +\alpha(b(x))^2/p$, $F_V(x)=\mathcal{L}_fV(x)+\mathcal{L}_gV(x)\pi(x)+\beta(b(x))\gamma(x)$. The critical regions that appear in \eqref{eq:controller} are defined by 
    \begin{subequations}\label{eq:criticalregion}
        \begin{align}
            &\Omega_{\overline{clf}}^{\overline{cbf}}=\{x\in\mathbb{R}^n:F_b>0,F_V<0\},\label{eq:cr1}\\
            &\Omega_{\overline{clf},1}^{cbf}=\{x\in\mathbb{R}^n:F_VF_b'-F_b\mathcal{L}_gV\mathcal{L}_gb^\top <0,\nonumber\\
            &~~~~~~~~~F_b\le 0,F'_b\ne 0\},\label{eq:cr2}\\
            &\Omega_{\overline{clf},2}^{cbf}=\{x\in\mathbb{R}^n:F_V<0,b(x)=0,\mathcal{L}_gb=0\},\label{eq:cr3}\\
            &\Omega_{clf,1}^{\overline{cbf}}=\{x\in\mathbb{R}^n:F_b\mathcal{L}_gV\mathcal{L}_gV^\top -F_V\mathcal{L}_gb\mathcal{L}_gV^\top>0,\nonumber\\
            &~~~~~~~~~~F_V\ge 0,\mathcal{L}_gV\ne 0\},\label{eq:cr4}\\
        &\Omega_{clf,2}^{\overline{cbf}}=\{x\in\mathbb{R}^n:F_b>0,\mathcal{L}_gV=0\},\label{eq:cr5}\\
        &\Omega_{clf,1}^{cbf}=\{x\in\mathbb{R}^n:\mathcal{L}_gV\ne 0,F_b\mathcal{L}_gV\mathcal{L}_gb^\top +F'_bF_v\ge 0\nonumber\\
        &F_b\mathcal{L}_gV\mathcal{L}_gV^\top +F_V\mathcal{L}_gb\mathcal{L}_gV^\top\ge 0,b(x)\ne 0\},\label{eq:cr6}\\
        &\Omega_{clf,2}^{cbf}=\{x\in\mathbb{R}^n:\mathcal{L}_gV\ne 0,F_V\le 0,F'_b=0,F_b=0\},\label{eq:cr7}\\
        &\Omega_{clf,3}^{cbf}=\{x\in\mathbb{R}^n:\mathcal{L}_gV\ne0,b(x)=0,\nonumber\\
        &~~~~~~F_V\le 0,\mathcal{L}_gb//\mathcal{L}_gV,F_V\mathcal{L}_gV=-F_b\mathcal{L}_gb\}.\label{eq:cr8}\\
        &\Omega_{clf,4}^{cbf}=\{x\in\mathbb{R}^n:\mathcal{L}_gV=0,F_b\le 0,F'_b\ne 0,F_V=0\}.\label{eq:cr9}\\
        &\Omega_{clf,5}^{cbf}=\{x\in\mathbb{R}^n:\mathcal{L}_gV=0,F_b'=0,F_b=0,F_V=0\}.\label{eq:cr10}
        \end{align}
    \end{subequations}
    For the ease of notation, we have dropped the dependency of $x$ for $F_b(x),F_V(x),\mathcal{L}_b(x),\mathcal{L}_V(x),F'_b(x)$. 
\end{thm}
\vspace{-3ex}
\begin{pf}
    Dualizing the control barrier function constraint $-\mathcal{L}_fb-\mathcal{L}_gbu-s\alpha(b(x))\le 0$ with a multiplier $\lambda_1\ge 0$, and the control Lyapunov function constraint $\mathcal{L}_fV+\mathcal{L}_gVu+\beta(b(x))\gamma(x)\le 0$ with a multiplier $\lambda_2\ge 0$, we obtain the corresponding Lagrangian 
    \begin{align*}
    L(u,s,\lambda_1,\lambda_2) &= \frac{1}{2}||u-\pi(x)||^2+\frac{p}{2}(s-1)^2\\
    &-\lambda_1(\mathcal{L}_fb+\mathcal{L}_gbu+s\alpha(b(x)))\\
    &+\lambda_2(\mathcal{L}_fV+\mathcal{L}_gVu+\beta(b(x))\gamma(x)). 
    \end{align*}
    The KKT conditions are given by
    \begin{subequations}\label{eq:KKT_ourfilter}
    \begin{align}
            \left.\frac{\partial L}{\partial u}\right|_{u=u^*}=u^*-\pi(x)-\lambda_1\mathcal{L}_gb^\top+\lambda_2\mathcal{L}_gV^\top&=0,\label{eq:KKT_ourfilter_a}\\
            \left.\frac{\partial L}{\partial s}\right|_{s=s^*}=p(s^*-1)-\lambda_1\alpha(b(x))&=0,\label{eq:KKT_ourfilter_b}\\
\lambda_1\left(\mathcal{L}_fb+\mathcal{L}_gbu^*+s^*\alpha(b(x))\right)&=0,\label{eq:KKT_ourfilter_c}\\
\lambda_2\left(\mathcal{L}_fV+\mathcal{L}_gVu^*+\beta(b(x))\gamma(x)\right)&=0.\label{eq:KKT_ourfilter_d}
    \end{align}
    \end{subequations}
    
We highlight here that, $u^*$, $s^*$, $\lambda_1$ and $\lambda_2$ are all functions of $x$. The dependency on $x$ is dropped to simplify notation.
By regarding $x$ as a parameter to the quadratic programming problem \eqref{eq:ourfilter}, the analytical solution can be evaluated by considering which constraints are active or inactive \cite{wang2022explicit,tan2021undesired,ames2016control}. We consider the following cases.

\emph{Case 1: Both the CBF and the CLF constraint are inactive.}

In this case, we have 
\begin{equation}\label{eq:case1}
    F_1(u^*,s^*)>0 \text{ and }F_2(u^*)<0.
\end{equation}
Then we have $\lambda_1=0$, and $\lambda_2=0$ from the complementary slackness conditions \eqref{eq:KKT_ourfilter_c} \eqref{eq:KKT_ourfilter_d}. By  \eqref{eq:KKT_ourfilter_a} and \eqref{eq:KKT_ourfilter_b}, we obtain that $u^*=\pi(x)$ and $s^*=1$, respectively. {This case happens for $x\in\Omega_{\overline{clf}}^{\overline{cbf}}\subset\mathbb{R}^n$}, where 
\begin{equation}
    \Omega_{\overline{clf}}^{\overline{cbf}}=\{x\in\mathbb{R}^n:F_1(u^*,s^*)>0,F_2(u^*)<0\}.
\end{equation}
{Substituting $u^*=\pi(x),s^*=1$ into $F_1(u^*,s^*),F_2(u^*)$, we thus obtain \eqref{eq:cr1}.}

\emph{Case 2: The CLF constraint is inactive and the CBF constraint is active.}

In this case, we have $F_1(u^*,s^*)=0,F_2(u^*)<0.$
Then we directly have $\lambda_2=0$ from \eqref{eq:KKT_ourfilter_d}. From \eqref{eq:KKT_ourfilter_a} we obtain
\begin{equation}\label{eq:case2uex}
u^*=\pi(x)+\lambda_1\mathcal{L}_gb^\top,
\end{equation} while from \eqref{eq:KKT_ourfilter_b} we obtain 
\begin{equation}\label{eq:case2sex}
    s^*=1+\frac{\lambda_1\alpha(b(x))}{p}.
\end{equation} We then consider the following two sub-cases.

1) $F'_b=\mathcal{L}_gb\mathcal{L}_gb^\top +\alpha(b(x))^2/p\ne 0$. Substituting \eqref{eq:case2uex} and \eqref{eq:case2sex} into $F_1(u^*,s^*)=0$, we get
\begin{equation}\label{eq:case2lambda}
\begin{split}
    \lambda_1 = -\frac{\mathcal{L}_gb\pi(x)+\mathcal{L}_fb+\alpha(b(x))}{\mathcal{L}_gb\mathcal{L}_gb^\top+\alpha(b(x))^2/p}=-\frac{F_{b}}{F'_b}.
\end{split}
\end{equation}
Substituting then \eqref{eq:case2lambda} into \eqref{eq:case2uex} and \eqref{eq:case2sex} we obtain
\begin{equation}\label{eq:case2u}
    u^*=\pi(x)-\frac{F_b}{F'_b}\mathcal{L}_gb^\top, s^*=1-\frac{F_b\alpha(b(x))}{pF'_b}.
\end{equation}
The critical region $\Omega_{\overline{clf},1}^{cbf}\subset\mathbb{R}^n$ is then defined by 
\begin{equation}\label{eq:case2case1}
\begin{split}
    \Omega_{\overline{clf},1}^{cbf}=&\{x\in\mathbb{R}^n:F_2(u^*)<0,\lambda_1\ge 0,F'_b\ne 0\}.
\end{split}
\end{equation} 
{Substituting \eqref{eq:case2u} into \eqref{eq:case2case1}, and since $F_b'$ is positive in this case, we obtain \eqref{eq:cr2}.}

2) $F'_b= 0$. This implies $\mathcal{L}_gb=0$ and $b(x)=0$. From \eqref{eq:case2uex} and \eqref{eq:case2sex}, we have that 
\begin{equation}\label{eq:case2case1u}
    u^*=\pi(x) \text{ and } s^*=1.
\end{equation}
For this case, $\lambda_1$ can be any arbitrary non-negative scalar. This is since the decision variables $u$ and $s$ do not appear in the CBF constraint. The dual function depends solely on $\lambda_2$. The critical region $\Omega_{\overline{clf},2}^{cbf}$ is thus
\begin{equation}\label{eq:case2case2}
    \Omega_{\overline{clf},2}^{cbf}=\{x\in\mathbb{R}^n:F_V<0,b(x)=0,\mathcal{L}_gb=0\},
\end{equation}
which establishes \eqref{eq:cr3}.

\emph{Case 3: The CLF constraint is active and the CBF constraint is inactive.}

For this case, we have $F_1(u^*,s^*)>0,F_2(u^*)=0.$
By \eqref{eq:KKT_ourfilter_c} we have that $\lambda_1=0$. Substituting this into \eqref{eq:KKT_ourfilter_b}, we obtain $s^*=1$, while 
substituting $\lambda_1=0$ into \eqref{eq:KKT_ourfilter_a} yields
\begin{equation}\label{eq:case3uex}
    u^*=\pi(x)-\lambda_2\mathcal{L}_gV^\top.
\end{equation}
Two further sub-cases are considered depending on the different values of $\mathcal{L}_gV$.

1) $\mathcal{L}_gV\ne 0$. Substituting \eqref{eq:case3uex} into $F_2(u^*)=0$ we have
\begin{equation}\label{eq:case3lambda2}
    \begin{split}
\lambda_2&=\frac{\mathcal{L}_fV+\mathcal{L}_gV\pi(x)+\beta(b(x))\gamma(x)}{\mathcal{L}_gV\mathcal{L}_gV^\top}=\frac{F_V}{\mathcal{L}_gV\mathcal{L}_gV^\top}.
    \end{split}
\end{equation}
By \eqref{eq:case3lambda2}, \eqref{eq:case3uex} results in
\begin{equation}\label{eq:case3case1u}
    u^*=\pi(x)-\frac{F_V}{\mathcal{L}_gV\mathcal{L}_gV^\top}\mathcal{L}_gV^\top.
\end{equation}
The critical region $\Omega_{clf,1}^{\overline{cbf}}$ is then defined by
\begin{equation}\label{eq:case3case1}
\begin{split}
    \Omega_{clf,1}^{\overline{cbf}}=\{x\in\mathbb{R}^n:F_1(u^*,s^*)>0,\lambda_2\ge 0,\mathcal{L}_gV\ne 0\}.
\end{split}
\end{equation}
{Substituting \eqref{eq:case3case1u} and $s^*=1$ into \eqref{eq:case3case1}, and since $\mathcal{L}_gV$ is positive in this case, we obtain \eqref{eq:cr4}.}

2) $\mathcal{L}_gV=0$. By \eqref{eq:case3uex} we have $u^*=\pi(x)$. For this case, $\lambda_2$ can be any non-negative scalar. The critical region is then defined by
\begin{equation}\label{eq:case3case2}
    \Omega_{clf,2}^{\overline{cbf}}=\{x\in\mathbb{R}^n:F_1(u^*,s^*)>0,\mathcal{L}_gV=0\}.
\end{equation}
{Substituting $u^*=\pi$ and $s^*=1$ into \eqref{eq:case3case2}, we obtain \eqref{eq:cr5}.}

\emph{Case 4: Both the CBF and CLF constraint are active.}
In this case, we have that 
\begin{equation}\label{eq:case4}
    F_1(u^*,s^*)=0 \text{ and } F_2(u^*)=0.
\end{equation}
By \eqref{eq:KKT_ourfilter_a} and \eqref{eq:KKT_ourfilter_b}, respectively, we obtain that
\begin{equation}\label{eq:case4uex}
    u^*=\pi(x)+\lambda_1\mathcal{L}_gb^\top -\lambda_2\mathcal{L}_gV^\top,
\end{equation}
\begin{equation}\label{eq:case4sex}
    s^*=1+\frac{\lambda_1\alpha(b(x))}{p}.
\end{equation}
Substituting \eqref{eq:case4uex}, and \eqref{eq:case4sex} into \eqref{eq:case4}, we obtain a linear equation
\begin{equation}\label{eq:case4lambda}
    \begin{bmatrix}
        \mathcal{L}_gb\mathcal{L}_gb^\top+\frac{\alpha(b(x))^2}{p} & -\mathcal{L}_gb\mathcal{L}_gV^\top \\
        \mathcal{L}_gV\mathcal{L}_gb^\top & -\mathcal{L}_gV\mathcal{L}_gV^\top 
    \end{bmatrix}\begin{bmatrix}
        \lambda_1\\\lambda_2
    \end{bmatrix}=\begin{bmatrix}
        -F_b(x)\\F_V(x)
    \end{bmatrix}
\end{equation}
Denote the matrix that pre-multiplies $[\lambda_1,\lambda_2]^\top$ by 
$A(x)$, and notice that its determinant is given by
\begin{align*}
\mathrm{det}&(A(x))=-\mathcal{L}_gV\mathcal{L}_gV^\top\mathcal{L}_gb\mathcal{L}_gb^\top \\
&-\mathcal{L}_gV\mathcal{L}_gV^\top \alpha(b(x))^2/p+\mathcal{L}_gb\mathcal{L}_gV^\top \mathcal{L}_gV\mathcal{L}_gb^\top. 
\end{align*}
It can be observed that $\mathrm{det}(A(x))=0$ if and only if (i) $\mathcal{L}_gV=0$; or, (ii) $F'_b(x)=\mathcal{L}_gb(x)\mathcal{L}_gb(x)^\top +\alpha(b(x))^2/p=0$, which in turn implies that $\mathcal{L}_gb=0$ and $b=0$; or, (iii) $b(x)=0$ and $\mathcal{L}_gV//\mathcal{L}_gb$. 

We then consider the following five sub-cases, that capture the situation where $\mathrm{det}(A(x)) \neq 0$, and all possible (distinct) cases for which $\mathrm{det}(A(x)) = 0$, and for each case characterize the resulting optimal solution of \eqref{eq:ourfilter}.

1) $\mathcal{L}_gV\ne  0$ and $b(x)\ne 0$. For this case we have $\det(A(x))\ne 0$. We then have that $A(x)$ is invertible with $A(x)^{-1}$ given by
    \begin{equation}\label{eq:ainverse}
        A(x)^{-1}=\frac{\begin{bmatrix}
            -\mathcal{L}_gV\mathcal{L}_gV^\top & \mathcal{L}_gb\mathcal{L}_gV^\top\\-\mathcal{L}_gV\mathcal{L}_gb^\top&\mathcal{L}_gb\mathcal{L}_gb^\top+\frac{\alpha(b(x))^2}{p}
        \end{bmatrix}}{\mathrm{det}(A(x))}.
    \end{equation}
We could thus solve the linear system of equations in \eqref{eq:case4lambda} and obtain
    \begin{equation}\label{eq:lineq}
        \begin{bmatrix}
            \lambda_1\\
            \lambda_2
        \end{bmatrix}=A(x)^{-1}\begin{bmatrix}
            -F_b(x)\\F_V(x)
        \end{bmatrix}.
    \end{equation}
    Substituting \eqref{eq:lineq}-\eqref{eq:ainverse} into \eqref{eq:case4uex} and \eqref{eq:case4sex}, we obtain the expressions for 
    $u^*$ and $s^*$ given in \eqref{eq:control4}. The critical region $\Omega_{clf,1}^{cbf}$ is defined by 
    \begin{equation}
\Omega_{clf,1}^{cbf}=\{x\in\mathbb{R}^n:\lambda_1\ge 0,\lambda_2\ge 0,\mathcal{L}_gV(x)\ne 0,b\ne 0\}. 
    \end{equation}
Under the expressions for $\lambda_1$ and $\lambda_2$ in \eqref{eq:lineq}, $A(x)^{-1}$ in \eqref{eq:ainverse}, and recalling that $F'_b:=\mathcal{L}_gb\mathcal{L}_gb^\top+\frac{\alpha(b(x))^2}{p}$, we obtain \eqref{eq:cr6}. 

2) $\mathcal{L}_gV\ne 0$ and $F'_b=0$. $F'_b=0$ implies that $\mathcal{L}_gb=0$ and $b=0$. Therefore, the linear system of equations in \eqref{eq:case4lambda} reduces to
\begin{equation*}
    \begin{split}
        0=-F_b \text{ and } -\mathcal{L}_gV\mathcal{L}_gV^\top \lambda_2=F_V.
    \end{split}
\end{equation*}
We then have $\lambda_2=-\frac{F_V}{\mathcal{L}_gV\mathcal{L}_gV^\top }$, and $F_b=\mathcal{L}_fb=0$, while $\lambda_1$ takes any arbitrary non-negative value. Substituting these identities into \eqref{eq:case4uex} and \eqref{eq:case4sex}, we obtain
\begin{equation}\label{eq:case4sc2u}
    u^*=\pi+\frac{F_V}{\mathcal{L}_gV\mathcal{L}_gV^\top}\mathcal{L}_gV^\top \text{ and }s^*=1.
\end{equation}
The critical region $\Omega_{clf,2}^{cbf}$ is then given by
\begin{equation}
\begin{split}
    \Omega_{clf,2}^{cbf}:=\{x\in\mathbb{R}^n:\lambda_2\ge 0,\mathcal{L}_gV\ne 0,F'_b=0,F_b=0\}.
    \end{split}
\end{equation}
Substituting in the latter $\lambda_2=-\frac{F_V}{\mathcal{L}_gV\mathcal{L}_gV^\top}$ yields \eqref{eq:cr7}.

3) $\mathcal{L}_gV\ne0, b(x)=0$, and $\mathcal{L}_gV//\mathcal{L}_gb$. For this case, matrix $A(x)$ has two linearly dependent rows. Therefore, the linear equation is feasible if and only if
\begin{equation*}
    F_V\mathcal{L}_gV=-F_b\mathcal{L}_gb.
\end{equation*}
Under this condition, the linear equation is feasible and admits infinitely many solutions. Fixing $\lambda_1=0$, by \eqref{eq:case4lambda} we have that $\lambda_2=-\frac{F_V}{\mathcal{L}_gV\mathcal{L}_gV^\top}$. Substituting these identities into \eqref{eq:case4uex} and \eqref{eq:case4sex}, we have
\begin{equation}\label{eq:case4sc3u}
    u^*=\pi+\frac{F_V}{\mathcal{L}_gV\mathcal{L}_gV^\top}\mathcal{L}_gV^\top \text{ and } s^*=1.
\end{equation}
The critical region $\Omega_{clf,3}^{cbf}$ is given by
\begin{equation}\label{eq:case4sc3cr}
\begin{split}
    &\Omega_{clf,3}^{cbf}:=\{x\in\mathbb{R}^n:\mathcal{L}_gV\ne 0,b(x)=0,\\
    &\lambda_2\ge 0,\mathcal{L}_gb//\mathcal{L}_gV,F_V\mathcal{L}_gV=-F_b\mathcal{L}_gb\}.
\end{split}
\end{equation}
Substituting $\lambda_2=-\frac{F_V}{\mathcal{L}_gV\mathcal{L}_gV^\top}$ into \eqref{eq:case4sc3cr}, we obtain \eqref{eq:cr8}.

4) $\mathcal{L}_gV=0$ and $F'_b\ne 0$. For this case, the linear equation \eqref{eq:lineq} results in
\begin{equation*}
    F'_b\lambda_1=-F_b,\quad0=F_V.
\end{equation*}
We thus have that $\lambda_1 = -\frac{F_b}{F'_b},$ while $\lambda_2$ takes any arbitrary non-negative value. Substituting these into \eqref{eq:case4uex} and \eqref{eq:case4sex}, we have 
\begin{equation}\label{eq:53}
    u^*=\pi-\frac{F_b}{F'_b}\mathcal{L}_gb^\top, \text{ and } s^*=1-\frac{F_b\alpha(b(x))}{pF'_b}.
\end{equation}
The critical region $\Omega_{clf,4}^{cbf}$ is given by
\begin{equation}\label{eq:case4sc4cr}
\begin{split}
    \Omega_{clf,4}^{cbf}:=&\{x\in\mathbb{R}^n:\lambda_1\ge 0,\mathcal{L}_gV=0,F'_b\ne 0,F_V=0\}.
\end{split}
\end{equation}
Substituting $\lambda_1=-\frac{F_b}{F_b'}$ into \eqref{eq:case4sc4cr}, we obtain \eqref{eq:cr9}.

5) $\mathcal{L}_gV=0$ and $F'_b= 0$. For this case, both $\lambda_1$ and $\lambda_2$ take arbitrary non-negative values. Besides, as a consequence of \eqref{eq:case4lambda}, we have $F_b=0,F_V=0$. By \eqref{eq:case4uex} and \eqref{eq:case4sex}, we then have 
\begin{equation}
    u^*=\pi(x),s^*=1.
\end{equation} The critical region $\Omega_{clf,5}^{cbf}$ is given by
\begin{equation*}
\begin{split}
\Omega_{clf,5}^{cbf}:=\{x\in\mathbb{R}^n:\mathcal{L}_gV=0,
   F'_b=0,F_b=0,F_V=0\},
\end{split}
\end{equation*}
which is \eqref{eq:cr10}, thus concluding the proof.\hfill\qed
\end{pf}
In the next theorem, we analyze the equilibrium points of our the closed-loop system $\dot x = f(x)+g(x)u^*(x)$, where $u^*(x)$ is obtained by solving \eqref{eq:ourfilter}.

\begin{thm}
    Consider system \eqref{eq:system}. Suppose there exist a CBF $b(x)$ and a CLF $V(x)$ that satisfy the relaxed compatibility condition. Let $u^*(x)$ be the optimal solution of \eqref{eq:ourfilter}. Then, the set of equilibrium points of system $\dot x = f(x)+g(x)u^*(x)$ in $\mathcal{B}$ are given by
    \begin{equation}\label{eq:region}
\mathcal{E}=\mathcal{E}_{cbf,1}^{clf}\cup\mathcal{E}_{cbf,2}^{clf}\cup\mathcal{E}_{cbf,3}^{clf}\cup\{0\}
    \end{equation}
    where
    \begin{subequations}\label{eq:regions}
        \begin{align}
            \mathcal{E}_{cbf,1}^{clf}&=\{x\in\{\Omega_{cbf,2}^{clf}\cup\Omega_{cbf,3}^{clf}\}\cap\partial \mathcal{B}:f(x)+g(x)\pi(x)\nonumber\\
            &+g(x)\frac{F_V}{\mathcal{L}_gV\mathcal{L}_gV^\top}\mathcal{L}_gV^\top =0\},\label{eq:region1}\\
             \mathcal{E}_{cbf,2}^{clf}&=\{x\in\{\Omega_{cbf,4}^{clf}\cap\partial \mathcal{B}\}:f(x)+g(x)\pi(x)\nonumber\\
            &-g(x)\frac{F_b}{\mathcal{L}_gb\mathcal{L}_gb^\top}\mathcal{L}_gb^\top=0\}.\label{eq:region3} \\
            \mathcal{E}_{cbf,3}^{clf}&=\{x\in\{\Omega_{cbf,5}^{clf}\cap\partial \mathcal{B}\}:f(x)+g(x)\pi(x)=0\},\label{eq:region2}
        \end{align}
    \end{subequations}
\end{thm}

\begin{pf}
We first show that $x=0$ is included in the set of equilibrium points. To this end, notice that $f(0)+g(0)\pi(0)=0$, as the optimal solution of \eqref{eq:ourfilter} at the origin is given by $u^*(0)=\pi(0)$, and $\pi$ is defined such that $\pi(0)=0$. Therefore, the origin is 
an equilibrium point. 

    Notice first that for any $x\ne0$ that satisfies $b(x)>0$, the CLF constraint in \eqref{eq:ourfilter} implies $\dot V(x)=\frac{\partial V(x)}{\partial x}(f(x)+g(x)u^*(x))<0$. Therefore, we have $f(x)+g(x)u^*(x)\ne 0$. This in turn implies that there are no equilibrium points in $\mathrm{Int}(\mathcal{B})$, and all equilibrium points have to belong to $\partial \mathcal{B}$. 
    
    We next investigate which constraints can be active for points on $\partial \mathcal{B}$. To this end, suppose that either the CBF constraint $\dot b(x)\ge 0$ or the CLF constraint $\dot V(x)\le 0$ is inactive. We have either $\dot b(x)=\frac{\partial b(x)}{x}(f(x)+g(x)u^*(x))>0$ or $\dot V(x)=\frac{\partial V(x)}{\partial x}(f(x)+g(x)u^*(x))<0$. It thus follows that $f(x)+g(x)u^*(x)\ne 0$ for either case. Therefore, we conclude that the CBF constraint and the CLF constraint have to be both active, and $b(x)=0$ at all equilibrium points. This corresponds to {the second, third, forth, and fifth sub-cases of} Case 4 {in the proof of} Theorem \ref{th:analytical}. We show next that all these cases are encompassed in the sets of \eqref{eq:region}.

    In particular:\\
    (i) For the second and third sub-case of Case 4 {in the proof of} Theorem \ref{th:analytical}, $x\in\Omega_{cbf,2}^{clf}\cup\Omega_{cbf,3}^{clf}$. By Theorem \ref{th:analytical}, we thus have that {$u^*=\pi+\frac{F_V}{\mathcal{L}_gV\mathcal{L}_gV^\top}\mathcal{L}_gV^\top$}. Under this choice of $u^*$ we obtain \eqref{eq:region1} for any $x$ such that  $f(x)+g(x)u^*(x)=0$. \\
    (ii) For the fourth sub-case of Case 4 {in the proof of} Theorem \ref{th:analytical}, we have that $x\in\Omega_{cbf,4}^{clf}$. We then have that $\mathcal{L}_gV=0,F_b'=0$. Recalling that $F_b'=\mathcal{L}_gb\mathcal{L}_gb^\top +\alpha(b)^2/p=\mathcal{L}_gb\mathcal{L}_gb^\top$, following \eqref{eq:control2}, the optimal controller $u^*(x)$ is given by $u^*(x)=\pi-\frac{F_b}{\mathcal{L}_gb\mathcal{L}_gb^\top}\mathcal{L}_gb^\top$. Under this choice of $u^*$ we obtain \eqref{eq:region3} for any $x$ such that  $f(x)+g(x)u^*(x)=0$. \\
    (iii) For the fifth sub-case of Case 4 {in the proof of} Theorem \ref{th:analytical}, we have that $x\in\Omega_{cbf,5}^{clf}$. 
    By Theorem \ref{th:analytical}, we then have $u^*=\pi(x)$. Under this choice of $u^*$ we obtain \eqref{eq:region2} for any $x$ such that  $f(x)+g(x)u^*(x)=0$. \hfill\qed
\end{pf}

\subsection{Continuity Analysis}
We now analyze $u^*(x)$, the optimal solution of \eqref{eq:ourfilter}, in terms of its continuity properties. We first provide two definitions that will be employed in the sequel and are widely encountered in the literature \cite{jankovic2018robust,tan2021undesired}.
\begin{defn}[\emph{Small Control Property}]\label{def:scp}
    A CLF $V(x)$ is said to have the \emph{Small Control Property} (SCP) if there exists a compact set $\Omega$ and a locally Lipschitz continuous control law $u_c(x):\mathcal{X}\to\mathbb{R}^m$, such that $0\in\Omega$, $\mathrm{Int}(\Omega)\ne \emptyset$, $\lim_{x\to 0}u_c(x)\to0$ and
    \begin{equation*}
        \mathcal{L}_fV(x)+\mathcal{L}_gV(x)u_c(x)+\gamma(x) \le 0,\forall x\in\Omega.
    \end{equation*}
\end{defn}

\begin{defn}[\emph{Strict Complementary Slackness}] The CLF constraint in \eqref{eq:ourfilter} is said to have \emph{Strict Complementary Slackness} (SCS) if
    \begin{align*}
\mathcal{L}_fV(x)+\gamma(x) <0,~\forall x\in\mathcal{X}\backslash\{0\} \text{ such that } \mathcal{L}_gV(x)=0.\label{eq:scsclf}
    \end{align*}
    
\end{defn}

SCS can be easily satisfied for CLFs. To see this, consider CLF $V(x)$, feedback controller $u(x)$, and function $\gamma(x)$ that satisfy \eqref{eq:clfflow}. Suppose there exists a set $\tilde{\mathcal{X}}\subseteq\{\mathcal{X}\cap\{x:\mathcal{L}_gV(x)=0\}\backslash\{0\}\}$, such that $\mathcal{L}_fV(x)+\gamma(x)=0,\forall x\in\tilde{\mathcal{X}}$. Define a new function $\tilde{\gamma}(x)=\gamma(x)/2$. Given that $0<\tilde{\gamma}(x)<\gamma(x),\forall x\in\tilde{\mathcal{X}}$, we deduce that $\mathcal{L}_fV(x)+\tilde{\gamma}(x)<0,\forall x\in\tilde{\mathcal{X}}$. Hence, the CLF constraint in \eqref{eq:ourfilter} can satisfy SCS by replacing $\gamma(x)$ with $\tilde{\gamma}(x)$.

\begin{thm}\label{th:continuity}
    {Consider system \eqref{eq:system}.} Suppose there exist CBF $b(x)$ and CLF $V(x)$ that satisfy the relaxed compatibility condition, and $0\in\mathrm{Int}(\mathcal{B})$. {Suppose $f(x)$, $g(x)$ and $\pi(x)$ are locally Lipschitz continuous on every compact subset of $\mathrm{Int}(\mathcal{B})\backslash\{0\}$.} Let $u^*(x)$ be the optimal controller obtained by \eqref{eq:ourfilter}. If SCS holds, then $u^*(x)$ is locally Lipschitz continuous $\forall x \in\mathrm{Int}(\mathcal{B})\backslash\{0\}$. Additionally, if SCP holds, then $u^*(x)$ is also locally Lipschitz continuous at 0.
\end{thm}
\begin{pf}
The proof follows a similar process as that for \cite[Theorem 1]{jankovic2018robust}.
To simplify notation, we let $\Omega_{\overline{clf}}^{cbf}=\Omega_{\overline{clf},1}^{cbf}\cup\Omega_{\overline{clf},2}^{cbf}$, $\Omega_{clf}^{\overline{cbf}}=\Omega_{clf,1}^{\overline{cbf}}\cup\Omega_{clf,2}^{\overline{cbf}}$, $\Omega_{clf}^{cbf}=\Omega_{clf,1}^{cbf}\cup\Omega_{clf,2}^{cbf}\cup\Omega_{clf,3}^{cbf}\cup\Omega_{clf,4}^{cbf}\cup\Omega_{clf,5}^{cbf}$. Therefore, the coefficient matrix $M(x)$ that multiplies $[u^\top,s]^\top$ in the active constraints of \eqref{eq:ourfilter} is given by

\begin{subequations}
    \begin{empheq}[left={M(x)=\empheqlbrace\,}]{align}
      &0,&\forall x\in\Omega_{\overline{clf}}^{\overline{cbf}},\\
      &\begin{bmatrix}
            \mathcal{L}_gV\quad0
        \end{bmatrix},&\forall x\in\Omega_{clf}^{\overline{cbf}},\\
        &\begin{bmatrix}
            \mathcal{L}_gb\quad\alpha(b(x))
        \end{bmatrix},&\forall x\in\Omega_{\overline{clf}}^{cbf},\\
        &\begin{bmatrix}
                \mathcal{L}_gb&\alpha(b(x))\\\mathcal{L}_gV&0
            \end{bmatrix},&\forall x\in\Omega_{clf}^{cbf}.
    \end{empheq}
\end{subequations}

Select a convex and compact set $\mathcal{D}\subset\mathrm{Int}(\mathcal{B})\backslash\{0\}$. The solution of \eqref{eq:ourfilter} in $\mathcal{D}$ is unique since the cost function is strictly convex and the constraints are linear. Given that $\mathcal{D}$ is compact, $\mathcal{L}_gb$, $\mathcal{L}_gV$, and $\alpha(b(x))$ are all {locally }Lipschitz continuous (since $f(x)$, $g(x)$, $V(x)$, $b(x)$, and $\alpha(\cdot)$ are all {locally }Lipschitz continuous), there exists $\zeta>0$ such that $||M(x)||<\zeta$ for every $x\in\mathcal{D}$. By affinity of the constraints this then implies that $u^*(x)$ is locally Lipschitz continuous for $x\in\mathrm{cl}(\Omega_{clf}^{cbf})\cap\mathcal{D}$ since $u^*(x)=\pi(x)$. If $\mathcal{L}_gV=0$ for some $x\in\Omega_{clf}^{\overline{cbf}}\cap\mathcal{D}$, the CLF constraint is necessarily inactive since $\mathcal{L}_fV+\beta(b(x))\gamma(x)\le \mathcal{L}_fV+\gamma(x)<0$ according to \eqref{eq:scsclf}, which contradicts the definition of $\Omega_{clf}^{\overline{cbf}}$ in \eqref{eq:criticalregion}. Therefore, for every $x\in\Omega_{clf}^{\overline{cbf}}\cap\mathcal{D}$, we have $\mathcal{L}_gV\ne0$. Similarly, for every $x\in\mathrm{cl}(\Omega_{\overline{clf}}^{cbf})\cap\mathcal{D}$, we have $\mathcal{L}_gV\ne 0,\alpha(b(x))\ne 0$. In summary, $M(x)$ is of full row rank in all cases. The conditions of \cite{jankovic2018robust} are satisfied, and the optimal solution $u^*(x)$ is locally Lipschitz continuous in every compact set $\mathcal{D}$. 

We then prove that if SCP holds, $u^*(x)$ is also {locally} Lipschitz continuous at the origin. Given that $0\in\mathrm{Int}(\mathcal{B})$ and $f(0)+g(0)\pi(0)=0$, we have $F_1(u_1(x),1)|_{x=0}=F_b(0)>0$ for any {locally} Lipschitz continuous control $u_1(x)$ that satisfies $u_1(0)=0$, which implies $0\in\mathrm{cl}(\Omega_{\overline{clf}}^{\overline{cbf}})\cup\mathrm{cl}(\Omega_{clf}^{\overline{cbf}})$. Therefore there exists a compact set $\mathcal{F}_{b}(u_1(x))\subset \mathrm{cl}(\Omega_{\overline{clf}}^{\overline{cbf}})\cup\mathrm{cl}(\Omega_{clf}^{\overline{cbf}})$ that satisfies $0\in\mathrm{Int}(\mathcal{F}_b(u_1(x)))$ and $F_1(u_1,1)>0~\forall x\in \mathcal{F}_{b}(u_1(x))$. From the SCP of $V(x)$, there exists a locally Lipschitz continuous control $u_2(x)$ that satisfies $u_2(0)=0$, and a compact set $\mathcal{F}_{V}(u_2(x))$ that satisfies $0\in\mathrm{Int}(\mathcal{F}_V(u_2(x)))$, such that $F_2(u_2(x)\le 0~\forall x\in\mathcal{F}_V(u_2(x))$. Define $\mathcal{F}(u_2(x))=\mathcal{F}_{V}(u_2(x))\cap\mathcal{F}_b(u_2(x))$. $\mathcal{F}(u_2(x))$ has a non-empty interior since $0\in\mathrm{Int}(\mathcal{F}_b(u_1(x)))$ and $0\in\mathrm{Int}(\mathcal{F}_V(u_2(x)))$. Then, for any $x\in\mathcal{F}(u_2(x))$,
\begin{align}
&F_2(u_2(x))=\mathcal{L}_fV+\mathcal{L}_gVu_2(x)+\beta(b(x))\gamma(x) \le 0\nonumber\\
&\Longleftrightarrow F_V+\mathcal{L}_gV(u_2(x)-\pi(x)) \le 0\nonumber\\
&\Longleftrightarrow |F_V|\le ||\mathcal{L}_gV||\cdot ||u_2(x)-\pi(x)||\nonumber\\
&\Longleftrightarrow \frac{|F_V|}{||\mathcal{L}_gV||}\le ||u_2(x)-\pi(x)||
\end{align}
Given that $\lim_{x\to 0}u_2(x)\to 0$, $\lim_{x\to 0}\pi(x)\to 0$, we have $\lim_{x\to 0}||u_2(x)-\pi(x)||\to 0$.
Hence, $\lim_{x\to 0}\frac{|F_V|}{||\mathcal{L}_gV||}\to 0$. Moreover, we have that the CBF constraint is inactive on $\mathcal{F}(u_2(x))$ since $\mathcal{F}(u_2(x))\subseteq \mathcal{F}_b(u_1(x))$. From \eqref{eq:control1} and \eqref{eq:control3}, we have $\lim_{x\to 0}u^*(x)\to 0$, for any $x\in\mathrm{cl}(\Omega_{\overline{clf}}^{\overline{cbf}})\cup\mathrm{cl}(\Omega_{clf}^{\overline{cbf}})$. Therefore, $u^*(x)$ is locally Lipschitz continuous on a compact set $\mathcal{F}(u^*)\subset \mathrm{cl}(\Omega_{\overline{clf}}^{\overline{cbf}})\cup\mathrm{cl}(\Omega_{clf}^{\overline{cbf}})$ that contains the origin. Hence, we conclude the proof.\hfill\qed
\end{pf}

Lipschitz continuity of $u^*(x)$, $f(x)$ and $g(x)$ guarantees uniqueness of solution for system \eqref{eq:system}. Using Theorem \ref{th:continuity}, we can then analyze the safety and stability performance of system \eqref{eq:system} using the optimal controller $u^*(x)$. 
\begin{lem}\label{lem:stability}
    Consider system \eqref{eq:system}. Suppose there exist a CBF $b(x)$ and a CLF $V(x)$ that satisfy the relaxed compatibility condition. Let $u^*(x)$ be the optimal controller obtained by solving \eqref{eq:ourfilter}. Then, for the closed-loop system $\dot x = f(x)+g(x)u^*(x)$ we have that:
    \begin{enumerate}
        \item The set $\mathcal{B}$ is forward invariant;
        \item The {CLF} $V(x)$ is decreasing for every $x\in\mathrm{Int}(\mathcal{B})\backslash\{0\}$.
    \end{enumerate}
\end{lem}
\begin{pf}
   Forward invariance is guaranteed since for any $x\in\partial \mathcal{B}$, $\dot b(x)=\mathcal{L}_fb+\mathcal{L}_gbu^*(x)\ge 0$, and $f(x)+g(x)u^*(x)$ is locally Lipschitz continuous. For every $x\in\mathrm{Int}(\mathcal{B})\backslash\{0\}$, the CLF $V(x)$ is decreasing since $\dot V=\mathcal{L}_fV+\mathcal{L}_gVu^*(x)< F_2(u^*(x))\le 0$.\hfill\qed
\end{pf}
\begin{thm}
    Consider system \eqref{eq:system}. Suppose there exist a CBF $b(x)$ and a CLF $V(x)$ that satisfy the relaxed compatibility condition following Definition \ref{def:compatible}. Let $u^*(x)$ be the optimal controller obtained from \eqref{eq:ourfilter}. If  
 $0\in\mathrm{Int}(\mathcal{B})$, then the closed-loop system \eqref{eq:system} $\dot x = f(x)+g(x)u^*(x)$ is locally asymptotically stable at the origin. Moreover, for any $l>0$, the set
    \begin{equation}\label{eq:roa}
        \mathcal{O}_l:=\{x\in\mathbb{R}^n:V(x)\le l\}
    \end{equation}
    is a region of attraction if $\mathcal{O}_l\cap\partial\mathcal{B}=\emptyset$.
\end{thm}
\begin{pf}
    For every $x\in\mathrm{Int}(\mathcal{B})\backslash\{0\}$, we have $\dot V(x)=\mathcal{L}_fV+\mathcal{L}_gVu^*(x)\le -\beta(b(x))\gamma(x)<0$. Any closed invariant set $\mathcal{R}$ that contains the origin is a region of attraction for the closed loop system. For any $l>0$ such that $\mathcal{O}_l\cap\partial\mathcal{B}=\emptyset$, we have $\dot V(x)\le 0$. Therefore, $\mathcal{O}_l$ is an invariant set for system $\dot x = f(x)+g(x)u^*$, and therefore a region of attraction.\hfill\qed
\end{pf}

\section{CLF \& CBF Design and Verification}
\label{sec:synthesis}
The proposed filter \eqref{eq:ourfilter} relies on pre-designed CLF $V(x)$ and CBF $b(x)$ that satisfy the relaxed compatibility condition. In this section, we show how to design these functions for a polynomial dynamical system and semi-algebraic sets $\mathcal{S}$, $\mathcal{X}$. After designing the functions offline, a controller that guarantees safety and local stability can be synthesized by solving \eqref{eq:ourfilter} in an online manner.
\begin{assum}\label{ass:polynomial}
    Consider system \eqref{eq:system}. $f(x)$ and $g(x)$ are polynomial functions. Assume that $\mathcal{S}$ and $\mathcal{X}$ are semi-algebraic sets, defined by $\mathcal{S}=\{x\in\mathbb{R}^n:s(x)\ge 0\}$, and $\mathcal{X}:=\{x\in\mathbb{R}^n:w(x)\ge 0\}$, respectively.
\end{assum}

To begin with the design method, we first restrict our decision variables such that $\sigma_1(x),\sigma_2(x),\sigma_3(x),\lambda_1(x),\\ \lambda_2(x),b(x),V(x),u_b(x),u_V(x)$. Additionally, we pre-define two sum-of-squares polynomials $\varepsilon_1(x)$ and $\varepsilon_2(x)$ with no constant terms. Clearly, we have
\begin{equation}\label{eq:varepsilon}
    \varepsilon_i(x)>0,\forall x\ne 0~\text{and}~\varepsilon_i(0)=0,\forall i\in\{1,2\}.
\end{equation}

We impose the following requirements on our design:\\
(i) Polynomial constraints: Using the previous identities, we require all decision variables to be polynomials. Namely,
\begin{equation}\label{eq:sospab}
\begin{split}
&\sigma_1(x),\sigma_2(x),\sigma_3(x)\in\Sigma[x],\\
&\lambda_1(x),\lambda_2(x),b(x),V(x),u_b(x),u_V(x)\in\mathbb{R}[x],b(0)>0.
\end{split}
\end{equation}
Specifically, $\{\sigma_i(x)\}_{i=1}^3$ are further restricted to be sum-of-squares polynomials, as they will be used to deduce set containment conditions using the first part of Lemma \ref{lem:S-procedure}. The value of CBF $b(x)$ is positive at $x=0$, which implies that $0\in\mathrm{Int}(\mathcal{B})$.

(ii) $\mathcal{B}\subseteq\mathcal{S}$: To enforce this, we introduce $\sigma_1(x)\in\Sigma[x]$ and use the first part of Lemma \ref{lem:S-procedure}. We then obtain
\begin{equation}\label{eq:sospc}
    -b(x)+\sigma_1(x)s(x)\in\Sigma[x].
\end{equation}

(iii) $\frac{\partial b(x)}{\partial x}(f(x)+g(x)u_b(x))\ge 0,\forall x\in\partial \mathcal{B}$: A sufficient condition to ensure this is by means of constraint \eqref{eq:sospd}. This involves introducing $\lambda_1(x)\in\mathbb{R}[x]$, and $u_b(x)\in\mathbb{R}[x]$, and considering the second part of Lemma \ref{lem:S-procedure}.
\begin{equation}\label{eq:sospd}
    \frac{\partial b(x)}{\partial x}(f(x)+g(x)u_b(x))+\lambda_1(x) b(x)\in\Sigma[x]
\end{equation}

(iv) $\frac{\partial V(x)}{\partial x}(f(x)+g(x)u_V(x))\le-\varepsilon_1(x)<0,\forall x\in\mathcal{X}\backslash\{0\}$: A sufficient condition to ensure this is by means of constraint \eqref{eq:sospe}. This involves introducing $u_V(x)\in\mathbb{R}[x]$, $\sigma_2(x)\in\Sigma[x]$, and considering \eqref{eq:varepsilon} and the first part of Lemma \ref{lem:S-procedure}.

\begin{equation}\label{eq:sospe}
    -\frac{\partial V(x)}{\partial x}(f(x)+g(x)u_V(x))-\sigma_2(x)w(x)-\varepsilon_1(x)\in\Sigma[x].
\end{equation}

(v) $V(x)\ge\varepsilon_2(x)>0,\forall x\in\mathcal{X}\backslash\{0\}$: A sufficient condition to ensure this is by means of constraint \eqref{eq:sospf}. This involves introducing $\sigma_3(x)\in\Sigma[x]$, and considering \eqref{eq:varepsilon} and the first part of Lemma \ref{lem:S-procedure}.

\begin{equation}\label{eq:sospf}
    V(x)-\varepsilon_2
(x)-\sigma_3(x)w(x)\in\Sigma[x].
\end{equation}

(vi) $\frac{\partial V(x)}{\partial x}(f(x)+g(x)u_b(x))\le 0,\forall x\in\partial \mathcal{B}$: A sufficient condition to ensure this is by means of constraint \eqref{eq:sospg}. This involves introducing $u_b(x)\in\mathbb{R}[x]$, and $\lambda_2(x)\in\mathbb{R}[x]$, and considering the second part of Lemma \ref{lem:S-procedure}. 
\begin{equation}\label{eq:sospg}
    -\frac{\partial V(x)}{\partial x}(f(x)+g(x)u_b(x))-\lambda_2(x)b(x)\in\Sigma[x].
\end{equation}

By incorporating all constraints in (i)-(vi) into one optimization (feasibility) problem, we propose the following program to design a CBF $b(x)$ and CLF $V(x)$ that satisfy the relaxed compatibility condition \eqref{eq:def3eq2}.

\begin{align}\label{eq:sosp}
    \mathrm{find}~&\{\sigma_i(x)\}_{i=1}^3,\{\lambda_i(x)\}_{i=1}^2,\nonumber \\
    &\hspace{2cm}b(x),V(x),u_b(x),u_V(x)\nonumber\\
    \mathrm{subject~to}~&\mathrm{constraints}~\eqref{eq:sospab}-\eqref{eq:sospg}.
\end{align}

The following theorem provides guarantees for the solution of \eqref{eq:sosp}.

\begin{thm}\label{th:sos}
    Consider Assumption \ref{ass:polynomial}, and further assume that a solution to \eqref{eq:sosp} exists and is denoted by $\{\sigma_i(x)\}_{i=1}^3,$ $\{\lambda_i(x)\}_{i=1}^2,$ $b(x),$ $V(x)$ $,u_b(x),$ $u_V(x)$. Then $b(x)$ is a CBF, $V(x)$ is a CLF, and they satisfy the relaxed compatibility condition as per Definition \ref{def:compalyapunov}.. 
\end{thm}
\begin{pf}
We will prove that: (i) if the solution satisfies constraints \eqref{eq:sospab}, \eqref{eq:sospc}, and \eqref{eq:sospd}, then $b(x)$ is a CBF; (ii) if the solution satisfies constraints \eqref{eq:sospab}, \eqref{eq:sospe}, and \eqref{eq:sospf}, then $V(x)$ is a CLF; (iii) if the solution satisfies constraints \eqref{eq:sospab}, \eqref{eq:sospd}, and \eqref{eq:sospg}, then $b(x)$ and $V(x)$ satisfy relaxed compatibility condition.
As such, if all constraints of \eqref{eq:sosp} are satisfied, all assertions of the theorem follow.

(i) Equation \eqref{eq:sospc} implies that for any $x\in\mathbb{R}^n$, $-b(x)+\sigma_1(x)s(x)\ge 0$, therefore for any $x\in\mathbb{R}^n$, $-b(x)+\sigma_1(x)s(x)>0$. From this, we have for any $x\in\mathbb{R}^n$ such that $s(x)<0$, we have that $b(x)<0$. This implies that $\mathcal{B}\subseteq \mathcal{S}$. Equation \eqref{eq:sospd} guarantees that for any $x$ such that $b(x)=0$, $\dot b(x)=\frac{\partial b(x)}{\partial x}(f(x)+g(x)u(x))\ge 0$. Thus, $b(x)$ is a CBF. 

(ii) Equation \eqref{eq:sospe} implies that for any $x\in\mathcal{X}$, $\dot V(x)+\varepsilon_1(x) \le 0$. \eqref{eq:sospf} indicates $V(x)-\varepsilon_2(x)\ge 0,\forall x\in\mathcal{X}$. Given that $\varepsilon_1(x)$ and $\varepsilon_2(x)$ satisfy \eqref{eq:varepsilon}, we hence conclude that $V(x)$ is a CLF. 

(iii) Equations \eqref{eq:sospd}, \eqref{eq:sospg} imply that $\mathcal{L}_fV+\mathcal{L}_gVu_b\le 0$, and $\mathcal{L}_fb+\mathcal{L}_gbu_b\ge 0$ for any $x$ such that $b(x)=0$. By Definition \ref{def:compalyapunov}, $b(x)$ and $V(x)$ satisfy the relaxed compatibility condition. \hfill\qed
\end{pf}

Naturally, the CLF $V(x)$ designed by \eqref{eq:sosp} will satisfy the SCP if $u_V(x)$ has no constant term. The CLF constraint in \eqref{eq:ourfilter} with $V(x)$ can satisfy the SCS by using $\gamma(x)=\varepsilon_1(x)/2$.

The program \eqref{eq:sosp} cannot be transformed into a semi-definite program due to the cross-product of the decision variables, e.g. $\frac{\partial b(x)}{\partial x}g(x)u_b(x)$. One tractable way to solve the problem is using an alternating directional algorithm, which solves the problem by alternating between the decision variables in iterations to handle bilinearities. The bilinearities in \eqref{eq:sosp} come from $\frac{\partial b(x)}{\partial x}u_b(x)$, and $\lambda_1(x)b(x)$ in \eqref{eq:sospd}; $\frac{\partial V(x)}{\partial x}u_V(x)$, and $\sigma_2(x)b(x)$ in \eqref{eq:sospe}; $\lambda_2(x)b(x)$, and $\frac{\partial V(x)}{\partial x}g(x)u_b(x)$ in \eqref{eq:sospg}. The decision variables can be separated into two groups: (i) $b(x)$ and $V(x)$; (ii) the others. If either group of variables is fixed, In the sequel, we will use the superscript $t$ to represent the corresponding fixed value of a decision variable at iteration $t$.





\begin{algorithm}[h]
 \caption{CLF and CBF design algorithm}
  \hspace*{\algorithmicindent} \textbf{Initialization} Functions $b^0(x)$ and $V^0(x)$, $t=1$.\\
 \hspace*{\algorithmicindent} \textbf{Output:} CBF $b(x)$ and CLF $V(x)$ that satisfy the relaxed compatibility condition \eqref{eq:def3eq2}\\
 \vspace{-3ex}
 \begin{algorithmic}[1]\label{al:dcbf}
 \WHILE{If \eqref{eq:sospiter} or \eqref{eq:sospiter2} is infeasible}
 \STATE \label{step:prog1} \textbf{Fix} $b^{t-1}(x)$ and $V^{t-1}(x)$. \\ \textbf{Solve} \eqref{eq:sospiter} for $\sigma_2^t(x)$, $\{\lambda_i^t(x)\}_{i=1}^2$, $u_b^t(x)$, and $u_V^t(x))$.

 \STATE \label{step:prog2} \textbf{Fix} $\{\lambda_i^t(x)\}_{i=1}^2$, $u_b^t(x)$, and $u_V^t(x)$. \\ \textbf{Solve} \eqref{eq:sospiter2} for $\{\sigma_{i}^t(x)\}_{i=1}^3$, $b^t(x)$, and $V^t(x)$.
 \ENDWHILE
 \end{algorithmic}
 \label{al:design}
\end{algorithm}
Here we propose Algorithm \ref{al:design} to design a CLF $V(x)$ and CBF $b(x)$ that satisfy the relaxed compatibility condition \ref{eq:def3eq2}. For initialization, we consider to design a valid local CLF $V^0(x)$ using the sum-of-squares techniques proposed by \cite{anderson2015advances}, and a valid CBF $b^0(x)$ using the methods proposed by \cite{wang2022safety}. In the algorithm, there are two main steps to iteratively solve two programs in each of which part of the variables in \eqref{eq:sosp} are fixed. At Step \ref{step:prog1}, we fix $b(x)$ and $V(x)$ by $b^{t-1}(x)$ and $V^{t-1}(x)$, respectively, to derive a convex program. For the first iteration $t=1$, $b^0(x)$ and $V^0(x)$ are obtained by the initialization step. The program is given by
\begin{subequations}\label{eq:sospiter}
    \begin{align}
    &\mathrm{find}~\sigma_2(x),\{\lambda_i(x)\}_{i=1}^2,u_b(x),u_V(x))\nonumber\\
&\sigma_2(x)\in\Sigma[x],\lambda_1(x),\lambda_2(x),u_b(x),u_V(x)\in\mathbb{R}[x],\label{eq:sospiterb}\\
    &\frac{\partial b^{t-1}(x)}{\partial x}(f(x)+g(x)u_b(x))+\lambda_1(x) b^{t-1}(x)\in\Sigma[x],\label{eq:sospiterc}\\
    &-\frac{\partial V^{t-1}(x)}{\partial x}(f(x)+g(x)u_V(x))-\nonumber\\
    &\sigma_2(x)w(x)-\varepsilon_1(x)\in\Sigma[x],\label{eq:sospiterd}\\
    &-\frac{\partial V^{t-1}(x)}{\partial x}(f(x)+g(x)u_b(x))-\lambda_2(x)b(x)\in\Sigma[x].\label{eq:sospiterf}
\end{align}
\end{subequations}

After solving \eqref{eq:sospiter} at Step \ref{step:prog1}, we fix $\{\lambda_i(x)\}_{i=1}^2$, $u_b(x)$ and $u_V(x)$ by $\{\lambda_i^t(x)\}_{i=1}^2$, $u_b^t(x)$ and $u_V^t(x)$, respectively, and solve another convex program at Step \ref{step:prog2}. The program is given by
\begin{subequations}\label{eq:sospiter2}
    \begin{align}
    &\mathrm{find}~\{\sigma_i(x)\}^3_{i=1},b(x),V(x)\\
&\sigma_1(x),\sigma_2(x),\sigma_3(x)\in\Sigma[x],b(x),V(x)\in\mathbb{R}[x],b(0)>0,\label{eq:sospiter2b}\\
    &-b(x)+\sigma_1(x)s(x)\in\Sigma[x],\label{eq:sospiter2c}\\
    &\frac{\partial b(x)}{\partial x}(f(x)+g(x)u_b^t(x))+\lambda_1^t(x) b(x)\in\Sigma[x],\label{eq:sospiter2d}\\
    &-\frac{\partial V(x)}{\partial x}(f(x)+g(x)u_V^t(x))-\sigma_2(x)w(x)-\varepsilon_1(x)\in\Sigma[x],\label{eq:sospiter2e}\\
    & V(x)-\varepsilon_2(x)-\sigma_3(x)w(x)\in\Sigma[x],\\
    &-\frac{\partial V(x)}{\partial x}(f(x)+g(x)u_b^t(x)-\lambda_2^t(x)b(x))\in\Sigma[x].\label{eq:sospiter2g}
\end{align}
\end{subequations}

Programs \eqref{eq:sospiter} and \eqref{eq:sospiter2} are both sum-of-squares programs; they are transformed into semi-definite programs and can be solved efficiently by an interior-point method. The algorithm terminates if both program \eqref{eq:sospiter} and \eqref{eq:sospiter2} are feasible across the same iteration.

Given a CBF $b(x)$ and a CLF $V(x)$, we can check the {relaxed} compatibility by a single SOS program:
\begin{subequations}\label{eq:sospcompatibility}
    \begin{align}
    &~~~~~~~~~~~~\mathrm{find}~\lambda_1(x),\lambda_2(x),u(x)~~~~~~~\mathrm{subject~to}~\nonumber\\
&\lambda_1(x),\lambda_2(x),u(x)\in\mathbb{R}[x],\label{eq:sospcompatibilitya}\\
    &\frac{\partial b(x)}{\partial x}(f(x)+g(x)u(x))+\lambda_1(x) b(x)\in\Sigma[x],\label{eq:sospcompatibilityb}\\
    &-\frac{\partial V(x)}{\partial x}(f(x)+g(x)u(x))-\lambda_2(x)b(x)\in\Sigma[x].\label{eq:sospcompatibilityc}
\end{align}
\end{subequations}

\section{Simulation Results}
\label{sec:simulation}
In this section, we demonstrate the performance of our designed filter \eqref{eq:ourfilter} by comparative studies over numerical examples. To solve the quadratic programming problem \eqref{eq:ourfilter} we leverage CVX \cite{grant2014cvx} for MATLAB. For the sum-of-squares programming problem we use SOSTOOLS v4.03 \cite{papachristodoulou2013sostools} to formulate the equivalent semi-definite programming problems, then solve these by using Mosek \cite{mosek}.

\subsection{Benchmark Case}
{Consider the same setting as Example \ref{ex:2}.} The term $\beta(b(x))$ is chosen to be $\beta(b(x))=\tanh(1000b(x))$. Relaxed compatibility of $b(x)$ and $V(x)$ can be validated by solving the proposed compatibility verification program \eqref{eq:sospcompatibility}. 

\begin{figure}[h]
    \centering
    \includegraphics[scale=0.6]{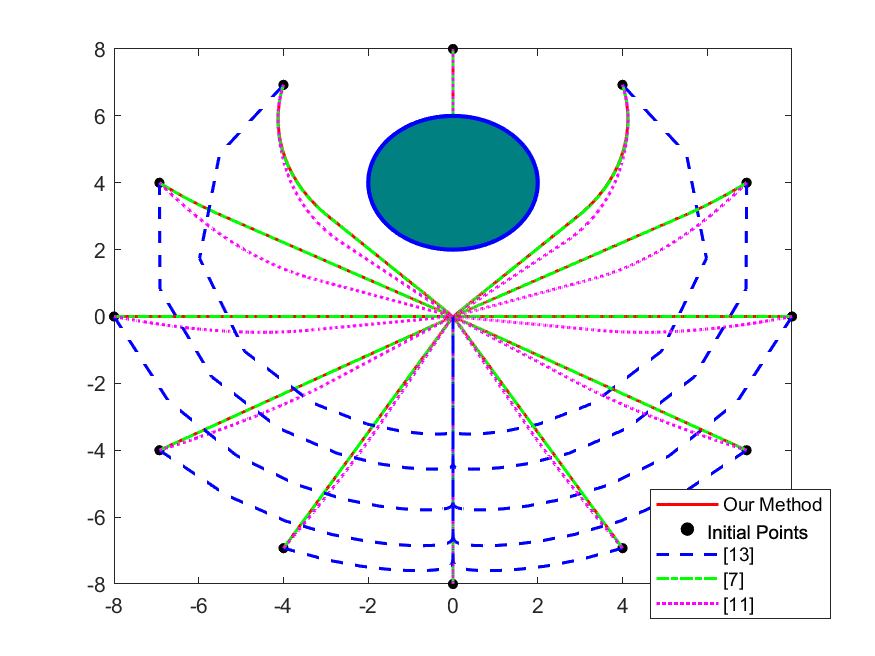}
    \caption{Trajectories of the closed-loop system with different methods. The black points are initial points, the green set is the obstacle. The penalty parameter $p_d$ is set to $100$ for the method of \cite{tan2021undesired}, \cite{ames2016control}, and our method. For the method of \cite{mestres2022optimization}, we set $\epsilon=0.01$ . Our method, \cite{tan2021undesired} and \cite{mestres2022optimization} achieve stabilization from all the initial points except for the top one, while \cite{ames2016control} only achieves inexact convergence. {From the very top initial point, all the trajectories converge to a point on the boundary of the obstacle. Safety is ensured by every method.}}
    \label{fig:case1fig1}
\end{figure}

\begin{figure}[h]
    \centering
    \includegraphics[scale=0.6]{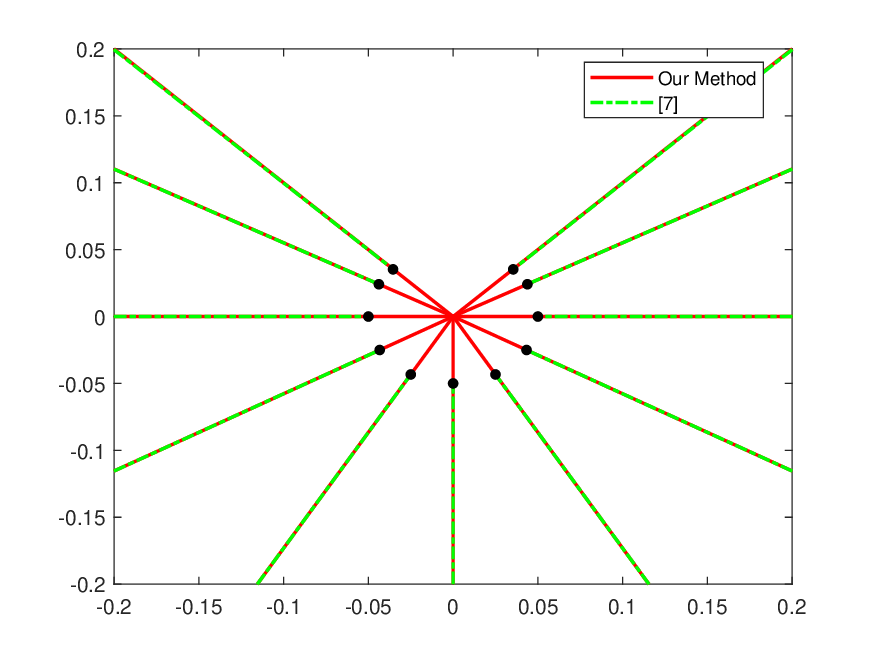}
    \caption{Trajectories near the origin, the red ones correspond to our method, while the green dotted ones correspond to the method of \cite{ames2016control}. All the trajectories of our methods converge to the origin, while these of \cite{ames2016control} converge to the black dots, which are equilibrium points away from the origin.}
    \label{fig:case1fig3}
\end{figure}

In Figure \ref{fig:case1fig1} we show trajectories from different initial points to compare our method \eqref{eq:ourfilter} with (i) the method of \cite{ames2016control}; (ii) the method of \cite{tan2021undesired}, 
and (iii) the method of  \cite{mestres2022optimization}. 
Trajectories generated by our method are almost aligned with these generated by the method of \cite{ames2016control}. 
The nominal controller $\pi(x)$ is set to zero to minimize the energy consumption in our method, the method of \cite{ames2016control} and \cite{mestres2022optimization}. To achieve local stability, a stabilizing nominal controller $\pi(x)=[-2x_1,-2x_2]^\top$ is chosen for the method of \cite{tan2021undesired}. The trajectories generated by our method are almost aligned with these by \cite{ames2016control}. 

In Figure \ref{fig:case1fig3}, we amplify the trajectories of our method and the methods of \cite{ames2016control} method near the origin. It can be observed that our method achieves convergence to the origin, whilst \cite{ames2016control} converges to equilibrium points away from the origin.

\begin{figure}[h]
    \centering
    \includegraphics[scale=0.6]{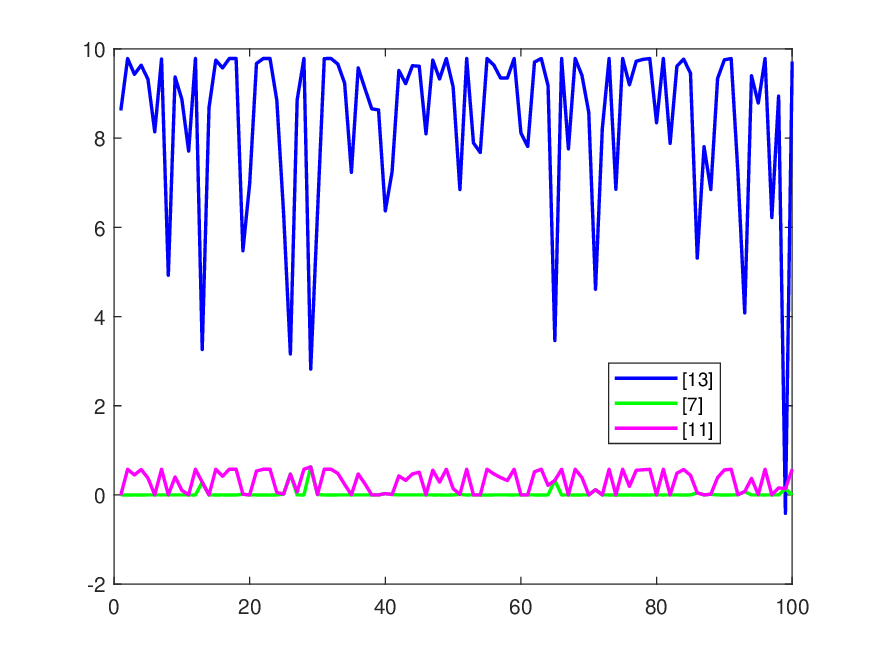}
    \caption{Comparison of filter performance by 100 Monte-Carlo experiments. The vertical axis represents $\log(||u_{\mathrm{o}}^*(x)||^2/||u^*(x)||^2)$, while the horizontal axis represents number of experiments.}
    \label{fig:case1fig2}
\end{figure}

We then compare the filter's performance for different methods. Performance is measured by the magnitude of $||u^*(x)||^2$. {Denote the optimal controller obtained from other methods by $u^*_{\mathrm{o}}$ (with $u^*_{\mathrm{o}}$ taking the value according to the different alternatives outlined above), and the one from our method by $u^*$. }We conduct 100 experiments, in each of which we randomly pick a point from $\mathcal{B}$, and calculate the value of the optimal controller obtained from each method. To enable a comparison, we set the performance of our filter $||u^*(x)||^2$ to be the base line, and plot the relative performance difference $\log(||u^*_{\mathrm{o}}(x)||^2/||u^*(x)||^2)$ in Figure \ref{fig:case1fig2}. It can be seen that our filter shows a better performance than the method of \cite{mestres2022optimization} and \cite{tan2021undesired}, and similar (better in some experiments) performance as the method of \cite{ames2016control}. Only in one experiment \cite{mestres2022optimization} exhibits better performance. We evaluate the controller $u^*(x)$ solved by the method of \cite{mestres2022optimization} at this point, find that $\mathcal{L}_fV(x)+\mathcal{L}_gV(x)u^*(x)+\gamma(x)> 0$. To have the CLF constraint satisfied for the method of \cite{mestres2022optimization} at this point, a smaller penalty coefficient $\epsilon$, \textit{i.e.,} $\epsilon=0.001$ is required. 

\subsection{Polynomial System}
Consider a second-order polynomial system with
\begin{equation}\label{eq:case2nonlin}
\begin{bmatrix}
    \dot x_1\\
    \dot x_2
\end{bmatrix}=
   \begin{bmatrix}
        x_2\\x_1+\frac{1}{3}x_1^3+x_2
    \end{bmatrix}+\begin{bmatrix}
        (0.2x_1^2+0.2x_2+1)u_1\\
        (-0.2x_2^2+0.2x_1+4)u_2
    \end{bmatrix}.
\end{equation}
The safe set is defined as $\mathcal{S}=\{x\in\mathbb{R}^2:x_1^2+(x_2-1)^2-0.25\}$. The local region is defined as $\mathcal{X}=-x_1^2-x_2^2+100$. We first verify the candidate CLF $V(x)=x^\top x$ for this system. We have $\mathcal{L}_fV(x)=4x_1x_2+\frac{2}{3}x_1^3x_2+2x_2^2$, $\mathcal{L}_gV(x)=[0.4x_1^3+0.4x_1x_2+2x_1,-0.4x_2^3+0.4x_1x_2+8x_2]^\top$. At state $x_1=0,x_2=\sqrt{20}$, we have $\mathcal{L}_gV(x)=0$, $\mathcal{L}_fV(x)=40>0$. By Definition \ref{def:CLF}, it follows that $V(x)$ is not a CLF.  


\begin{figure}[t]
    \centering
    \includegraphics[scale=0.6]{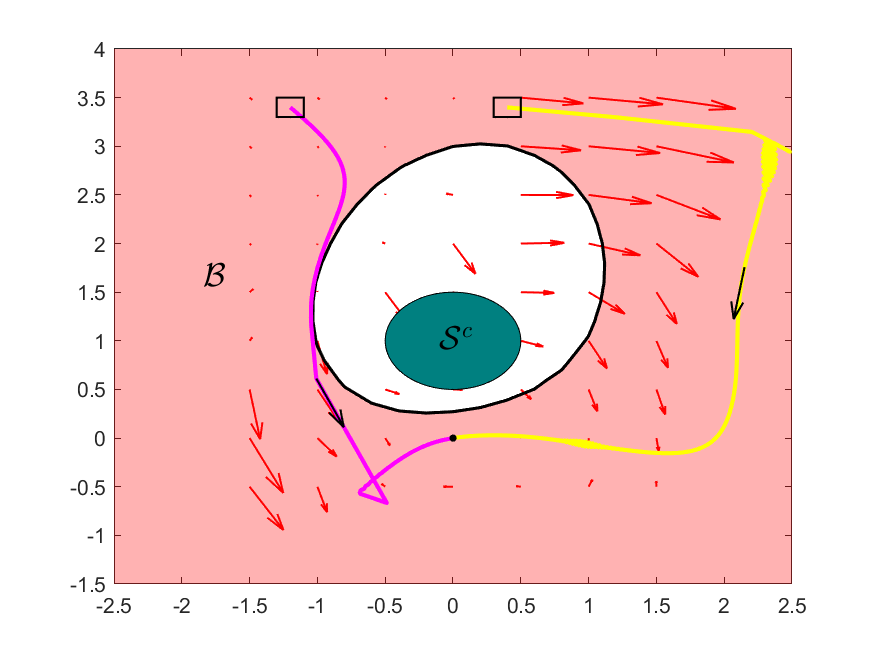}
    \caption{Phase portrait of system \eqref{eq:case2nonlin} using a CLF $V(x)$ and a CBF $b(x)$ that satisfy the \emph{relaxed} compatibility condition. The controller $u^*(x)$ is synthesized by solving \eqref{eq:ourfilter}, using the designed CLF and CBF. The green set represents the obstacle. The control invariant set $\mathcal{B}$ is filled in red while its boundary curve $\partial \mathcal{B}$ is highlighted in black. The red arrows represent the vector field $f(x)+g(x)u^*(x)$. Two trajectories start from the black rectangle, avoid $\mathcal{B}^c$, and finally converge to the origin.}
    \label{fig:case2fig1}
\end{figure}

\begin{table}[ht]
\centering
\caption{Degree of polynomial variables in \eqref{eq:sosp}.}
\begin{tabular}{|c|c|c|c|c|}
\hline
$\sigma_1(x)$ & $\lambda_1(x)$ & $\lambda_2(x)$ & $\sigma_2(x)$ & $\sigma_3(x)$ \\ \hline
2             & 1              & 4              & 8             & 8             \\ \hline \hline 
$b(x)$        & $V(x)$         & $u_b(x)$       & $u_V(x)$      &               \\ \hline
4             & 10             & 8              & 8             &               \\ \hline
\end{tabular}
\label{tab:1}
\end{table}

Using the Algorithm \ref{al:design}, a CBF $b(x)$ and a CLF $V(x)$ that satisfy the relaxed compatibility condition \ref{eq:def3eq2} can be designed. Degrees of polynomials in \eqref{eq:sosp} are shown in Table \ref{tab:1}. The sum-of-squares polynomial $\varepsilon_1(x)=\varepsilon_2(x)=0.1(x_1^2+x_2^2)$. The control invariant set $\mathcal{B}$ and level sets of $V(x)$ are shown in Figure \ref{fig:case2fig1}. It can be observed that the complementary set $\mathcal{B}^c$ is bounded that contains the obstacle, compatibility can not hold for this $b(x)$ and $V(x)$. The red arrows in the figure represent vector field $f(x)+g(x)u^*(x)$. On $\partial\mathcal{B}$, the arrows all point inwards $\mathcal{B}$, which reveals control invariance.

\begin{figure}
    \centering
    \includegraphics[scale=0.6]{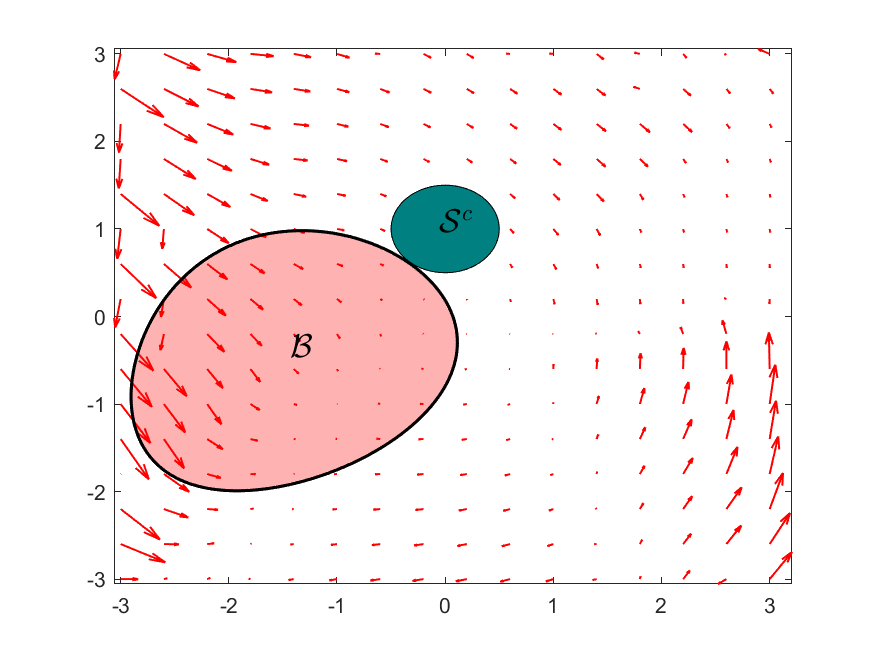}
    \caption{Phase portrait of system \eqref{eq:case2nonlin} using a CLF $V(x)$ and a CBF $b(x)$ that satisfy the \emph{strict} compatibility condition (Definition \ref{def:compatible}), designed by the algorithm proposed in \cite{schneeberger2024advanced}. The green set represents the obstacle, while the control invariant set $\mathcal{B}$ is filled in red. The red arrows represent the vector field $f(x)+g(x)u^*(x)$.}
    \label{fig:case2fig3}
\end{figure}

We then compare our method with \cite{schneeberger2024advanced}, which proposes a SOS program to design $b(x)$ and $V(x)$ that satisfy the strict compatibility condition, following Defition \ref{def:compatible}. Here we use the same degrees of polynomials for $b(x)$ and $V(x)$, and still consider $\varepsilon_1(x)=\varepsilon_2(x)=0.1(x_1^2+x_2^2)$. Using the new $b(x)$, set $\mathcal{B}$ is shown in Figure \ref{fig:case2fig3}. It can be seen that $\mathcal{B}$ is a bounded set, while our method generates an unbounded control invariant set. For this case, our method is shown to guarantee safety in a much larger region. 

\section{Conclusion}
\label{sec:conclusion}
In this paper, we have proposed a novel filter to design a safe and stable controller given a locally Lipschitz continuous reference signal. The filter is guaranteed to be feasible if the CBF and the CLF satisfy a relaxed compatibility condition. We have shown that the closed-loop system is safe, and locally stable at the origin. Any level set of the CLF that does not intersect with the control invariant set is guaranteed to be a region of attraction. By characterizing the closed-form solution of the filter, we have shown that there are no interior equilibrium points except for the origin. Moreover, we show that the designed optimal controller is locally Lipschitz continuous under mild regularity conditions. To obtain a CLF and CBF that satisfy the relaxed compatibility condition, we developed a sum-of-squares program for nonlinear polynomial dynamics and a semi-algebraic safe set. Future work concentrates on analyzing the stability properties of the boundary equilibrium points, and considering control input saturation constraints.
\bibliographystyle{ieeetr}
\bibliography{ref}
\end{document}